\documentclass[pre,preprint,showpacs,showkeys,preprintnumbers,amsmath,amssymb,floatfix]{revtex4}

\usepackage{amsmath}
\usepackage{bm}
\usepackage{amssymb}
\usepackage{graphics}
\usepackage{hyperref}
\usepackage{graphicx}
\usepackage{dcolumn}
\usepackage{bm}
\usepackage{subfigure}
\usepackage{threeparttable}

\newcommand{\thickhline}{\noalign{\hrule height 0.8pt}}

\begin{document}


\title{Dynamic phase transition properties of kinetic Ising model in the presence of additive white noise}
\author{Yusuf Y\"{u}ksel}
\email{yusuf.yuksel@deu.edu.tr}

\affiliation{Department of Physics, Dokuz Eyl\"{u}l University,
Kaynaklar Campus, TR-35160 Izmir, Turkey}
\date{\today}
\begin{abstract}
Using Monte Carlo simulations based on the Metropolis algorithm, we investigate the dynamic phase transition properties of kinetic Ising model 
driven by a sinusoidally oscillating magnetic field in the presence of additive white noise.
We calculate equilibrium and  dynamic  properties such as the temperature dependence of average magnetization and magnetic specific heat, as well as the period dependence of dynamic order parameter and scaled variance. 
After determining the critical period at which order-disorder transition takes place, we perform finite size scaling analysis to extract the exponent ratios, and discuss the variation
of these properties in the presence of noisy magnetic field. As a general result, we show that for a noisy system, DPT does not fall
into a universality class of the conventional dynamic (and also equilibrium) universality class of the Ising model.

\end{abstract}

\maketitle

\section{Introduction}\label{intro}
Dynamic phase transition (DPT) properties of kinetic Ising model have been investigated in detail during the past three decades \cite{tome,lo,chakrabarti}. Up to now, comprehensive theoretical efforts 
clarified several aspects of DPT. For instance, fluctuations of dynamic order parameter and energy in kinetic Ising model have been found to exhibit a singularity behavior at the transition temperature in the presence of 
oscillating magnetic fields \cite{acharyya1,acharyya2,acharyya3}. For large field amplitude and small lattice size, the system may exhibit a stochastic resonance behavior \cite{sides1,sides2} accompanied with 
a discontinuous transition observed in the thermal variation of the order parameter \cite{acharyya7}.
Besides, the studies regarding the dynamic hysteresis process of kinetic Ising model \cite{chakrabarti,acharyya5} revealed that a dynamic symmetry breaking 
 originates as a consequence of a competition between two time scales namely, the period of the dynamic order parameter and the relaxation time of the system resulting in a DPT between dynamically ordered and 
dynamically disordered phases. For a given field period $P$, the relaxation time depends on the several system parameters such as the temperature $T$, the magnetic field amplitude $h_{0}$, and the exchange interactions acting 
in the system. In several works, modified versions of the problem such as the kinetic Ising model with next-nearest neighbor interactions \cite{baez}, kinetic Blume-Capel (BC) model \cite{vatansever2},  
as well as site and bond diluted systems \cite{chattopadhyay,fytas} have been handled. 

On the other hand, for a $\mathrm{Co(4\AA)/Pt(7\AA)}$   multilayer system with strong perpendicular anisotropy, an example of DPT has been observed by Robb et al. \cite{robb}. Besides, 
very recently, DPT has also been experimentally observed for  uniaxial ferromagnetic films by Berger et al. \cite{berger} and  Riego et al. \cite{riego}. The experimental results reported in Ref. \cite{riego} 
have also been verified by Ref.\cite{buendia2} using numerical and theoretical tools. 
Moreover, regarding the universality class of the DPT observed in kinetic Ising model, two (2D) \cite{sides3,buendia,vatansever,robb2,korniss,sides5,sides4} and three (3D) \cite{park,tauscher} dimensional models have been widely investigated, and
it has been concluded that the critical exponents of the kinetic Ising system belongs to the same universality class of the equilibrium model. 

Although the kinetic Ising model has been widely investigated in the literature, very few works have paid attention on the  effect of randomness on the DPT properties of the model. Among them, 
Acharyya \cite{acharyya6} studied the kinetic Ising model in the presence of a randomly varying (in time but uniform in space) magnetic field. For this system, ordered phase disappears with increasing randomness
whereas for small random fields, the system remains in a dynamic asymmetric (ordered) phase. 
In the presence of a rectangular random field distribution, the same author reported the existence of a tricritical point at $T=0$ \cite{acharyya4}.
Haussman and Ruj\'{a}n \cite{haussman} investigated a kinetic Ising ferromagnet in the presence of a fast switching, random external field
which was realized according to a bimodal type of random field distribution. They found a novel type of first order phase transition in their system which was related to dynamic freezing. 
The fourth order cumulant (i.e. the Binder cumulant) and  dynamic magnetization-reversal transition analysis of kinetic Ising system in the presence of pulsed magnetic fields have been studied by Chatterjee and Chakrabarti \cite{chatterjee}
where they found that the transition to fall in a mean-field-like universality class. Meanwhile, Crokidakis \cite{crokidakis,crokidakis2} studied the random field kinetic Ising model 
where the local magnetic fields change sign randomly with time where the type of the random field is selected to be a double Gaussian type. According to his results, 2D system exhibits continuous (discontinuous) phase transition
for a bimodal (double Gaussian) random field whereas the transition for the 3D system becomes discontinuous for a bimodal field distribution with large field values.

Apart from these, very recently Akinci \cite{akinci} modeled a kinetic Ising system in the presence of both periodic and randomly fluctuating magnetic fields. As a source of randomness, the author considered the Gaussian white noise.
Effect of the white noise on the DPT properties of the system has been investigated, and it has been shown that there exists a white noise induced DPT in the system. 
Indeed, we encounter the noise (such as paper crumpling) in our every day life, mostly as an unwanted effect. It can also be observed in several ubiquitous systems including rotary 
motion of biological bacteria populations \cite{korobkova,tu,korobkova2}, voltage fluctuations across the resistance of electrical circuits \cite{weissman}, seismic activity during
earthquakes \cite{houston}, and  magnetization flips \cite{chen}, as well as random motion of magnetic domain walls \cite{benitez} of magnetic systems.
Depending on the nature of the noise (additive or multiplicative), effect of the noise on a particular property of a system may not always be of destructive kind but it may also play a constructive role.
Such behavior manifests itself as noise induced ordering and disordering transitions and reentrant phenomena in stochastic systems \cite{genovese,kim,anteneodo, xi,broeck,kharchenko,ojalvo,ojalvo2,li}.
To the best of our knowledge, there does not exist any work regarding the DPT properties of kinetic Ising model in the presence of noise. Therefore, it would be interesting to pursuit the answer 
for the question if there is any dynamic phase transition in the kinetic Ising system in the  presence of both periodic and noisy magnetic fields. For this aim, in addition to the conventional time dependent 
periodic part of the external magnetic field, we also consider the additive white noise acting on the lattice sites of the system.  
For this aim the paper is organized as follows: In Sec. \ref{model}, we present out model. The results and related discussions are given in Sec. \ref{results}. Finally, Sec. \ref{conclude} is devoted to our concluding remarks.

\section{Model and Simulation Details}\label{model}
In this work, we consider a kinetic Ising model defined on a square lattice with lattice coordination number $q=4$. The following Hamiltonian defines the dynamic behavior of the system
\begin{equation}\label{eq1}
\mathcal{H}=-J\sum_{<i,j>}S_{i}S_{j}-h(t)\sum_{i}S_{i}, 
\end{equation}
where the first term denotes the ferromagnetic $(J>0)$ exchange between nearest neighbor spins and the last summation which stands for the Zeeman energy contribution is carried over all the lattice sites. 
Within the framework of the Monte Carlo simulation method, the system defined by Eq. (\ref{eq1}) can be handled by several local spin update schemes such as 
\begin{eqnarray}\label{eq4}
\nonumber
W(S_{i}\rightarrow -S_{i})&=&\frac{\exp(-\beta\Delta E_{T})}{1+\exp(-\beta \Delta E_{T})}, \quad \mathrm{(Glauber)}\\ 
\nonumber
W(S_{i}\rightarrow -S_{i})&=&\frac{1}{1+\exp(\beta\Delta E_{J})}\frac{1}{1+\exp(\beta\Delta E_{H})}, \quad \mathrm{(Soft \ Glauber)}\\
W(S_{i}\rightarrow -S_{i})&=&\mathrm{Min}\left[1,\exp(-\beta\Delta E_{T})\right]. \quad \mathrm{(Metropolis)}
\end{eqnarray}
where $\Delta E_{T}=\Delta E_{J}+\Delta E_{H}$ is the local energy variation after flipping the spin $S_{i}$ at the site $i$.
In the present work, Eq. (\ref{eq1}) has been treated by Metropolis Monte Carlo scheme \cite{metropolis} with periodic boundary conditions on a $L\times L$ square lattice.
The time dependent magnetic field 
acting on the site $i$ is given by 
\begin{equation}\label{eq2}
h(t)=h_{0}\cos(\omega t)+h_{r}, 
\end{equation}
where $h_{0}$ and $\omega$ respectively correspond to the amplitude and the angular frequency of the periodic magnetic field, and $h_{r}$ is a random magnetic field which can be called as the additive Gaussian white noise.
At each time step of the simulation, a random value $h_{r}$ is drawn from the following normal probability distribution
\begin{equation}\label{eq3}
P(h_{r})=\frac{1}{\sqrt{2\pi\sigma^{2}}}\exp(-h_{r}^{2}/2\sigma^{2}), 
\end{equation}
where $\sigma$ is the width of the random field distribution. 
The simulation procedure can be outlined as follows: At first, we start by a random configuration of spins, and randomly visit the lattice sites (random sweeping) in one Monte Carlo
step per spin (MCSS). The period $P=2\pi/\omega$ of the periodic part of the external field is defined in terms of MCSS.
For the thermal variation of magnetic properties, the simulations run over 2000 periods of the external field cycle where $50\%$ of them were discarded for thermalization. On the other hand, for the calculations performed at 
constant temperature, our simulations have been carried out over $2\times10^{5}$  field periods where $10\%$ of them have been discarded to allow the system to reach equilibrium. 
Once the value of $h_{r}$ is assigned at a given simulation time, each spin on the lattice interacts with this noisy magnetic field during one complete sweep of the lattice sites
in one MCSS. During the simulation for a particular process, we have monitored a variety of quantities such as 
\begin{itemize}
\item the time series of the magnetization
\begin{equation}\label{eq5}
m(t)=\frac{1}{L^{2}}\sum_{i=1}^{L^{2}}S_{i},
\end{equation}
\item the averaged magnetization $M(T)$ and specific heat $C(T)$,
\begin{equation}\label{eq6}
M(T)=\left\langle m(t) \right\rangle, \quad C(T)=\frac{\partial \langle\mathcal{H}\rangle}{\partial T}, 
\end{equation}
 \item dynamic order parameter which can be defined as time averaged magnetization over the $k^{th}$ cycle of the periodic magnetic field,
 \begin{equation}\label{eq7}
  Q_{k}=\frac{1}{P}\int_{(k-1)P}^{kP}m(t)dt,
 \end{equation}
\item period average of the dynamic order parameter
\begin{equation}\label{eq8}
\langle |Q|\rangle_{L}=\frac{1}{N_{k}}\sum_{k=1}^{N_{k}}|Q_{k}|, 
\end{equation}
where $N_{k}$ is the number of the complete cycles of the periodic field.
\item We have also calculated the scaled variance, and the  fourth order cumulant (i.e. Binder cumulant) corresponding to the dynamic order parameter $Q$ which are respectively defined as
\begin{equation}\label{eq9}
\chi_{L}^{Q}=L^{2}[\langle Q^{2}\rangle_{L}-\langle |Q|\rangle^{2}_{L}], 
\end{equation}
\begin{equation}\label{eq10}
 U_{L}=1-\frac{\langle Q^{4}\rangle_{L}}{3\langle Q^{2}\rangle^{2}_{L}}.
\end{equation}
\end{itemize}
We have calculated the relations given between Eqs. (\ref{eq2})-(\ref{eq10}) for several lattice sites ranging from $L=64$ to $L=256$. The fixed temperature simulations have been performed at a temperature $T=0.8T_{c}$ where $T_{c}\approx2.269$
is the transition temperature of the equilibrium 2D Ising model on a square lattice. In order to investigate the DPT properties, we have also fixed the field amplitude as $h_{0}=0.3J$ for convention \cite{sides1}.
We also set $k_{B}=1$ throughout the work.

\section{Results and Discussion}\label{results}
\begin{figure}[!h]
\center
\subfigure[\hspace{0cm}] {\includegraphics[width=6.5cm]{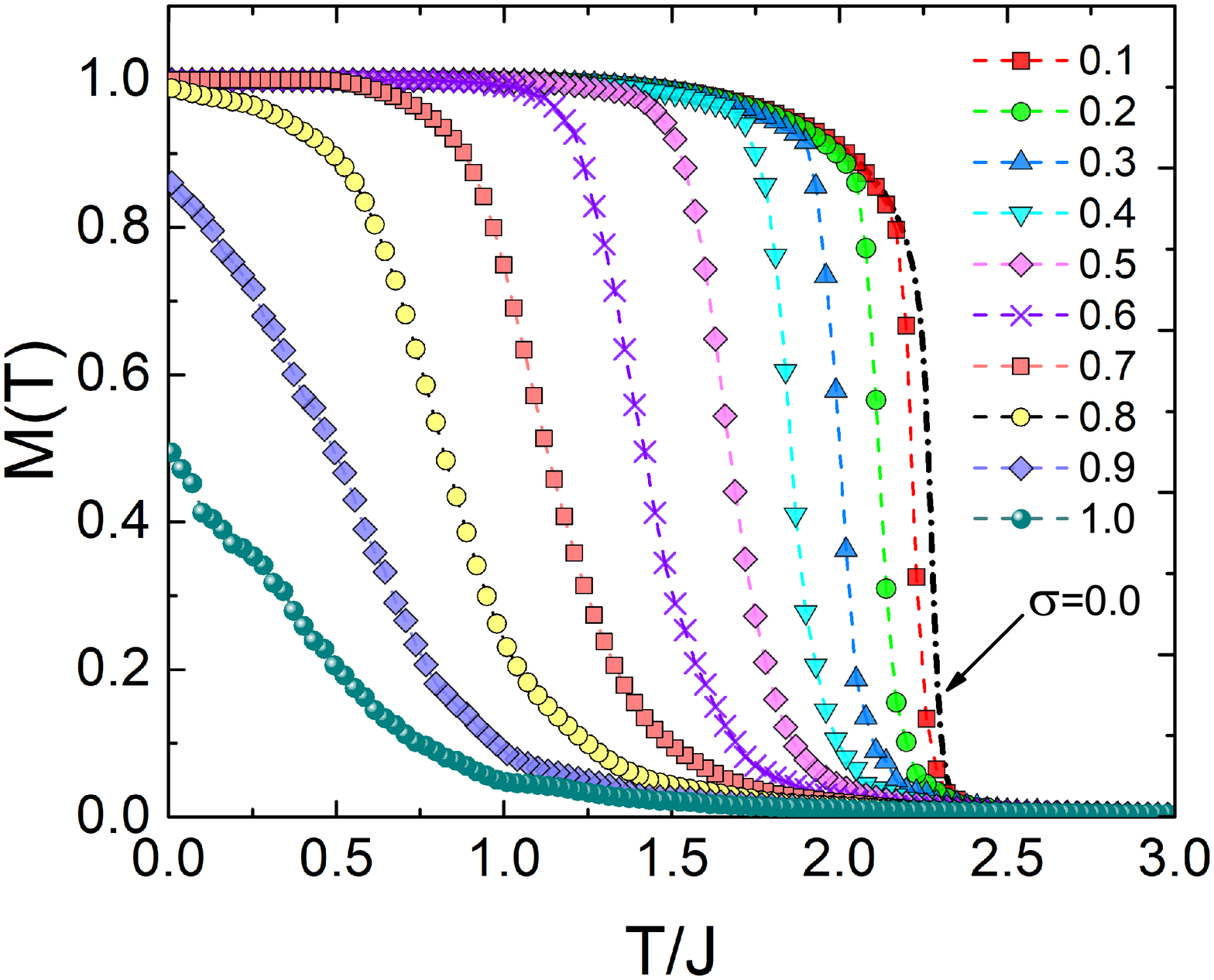}}
\subfigure[\hspace{0cm}] {\includegraphics[width=6.5cm]{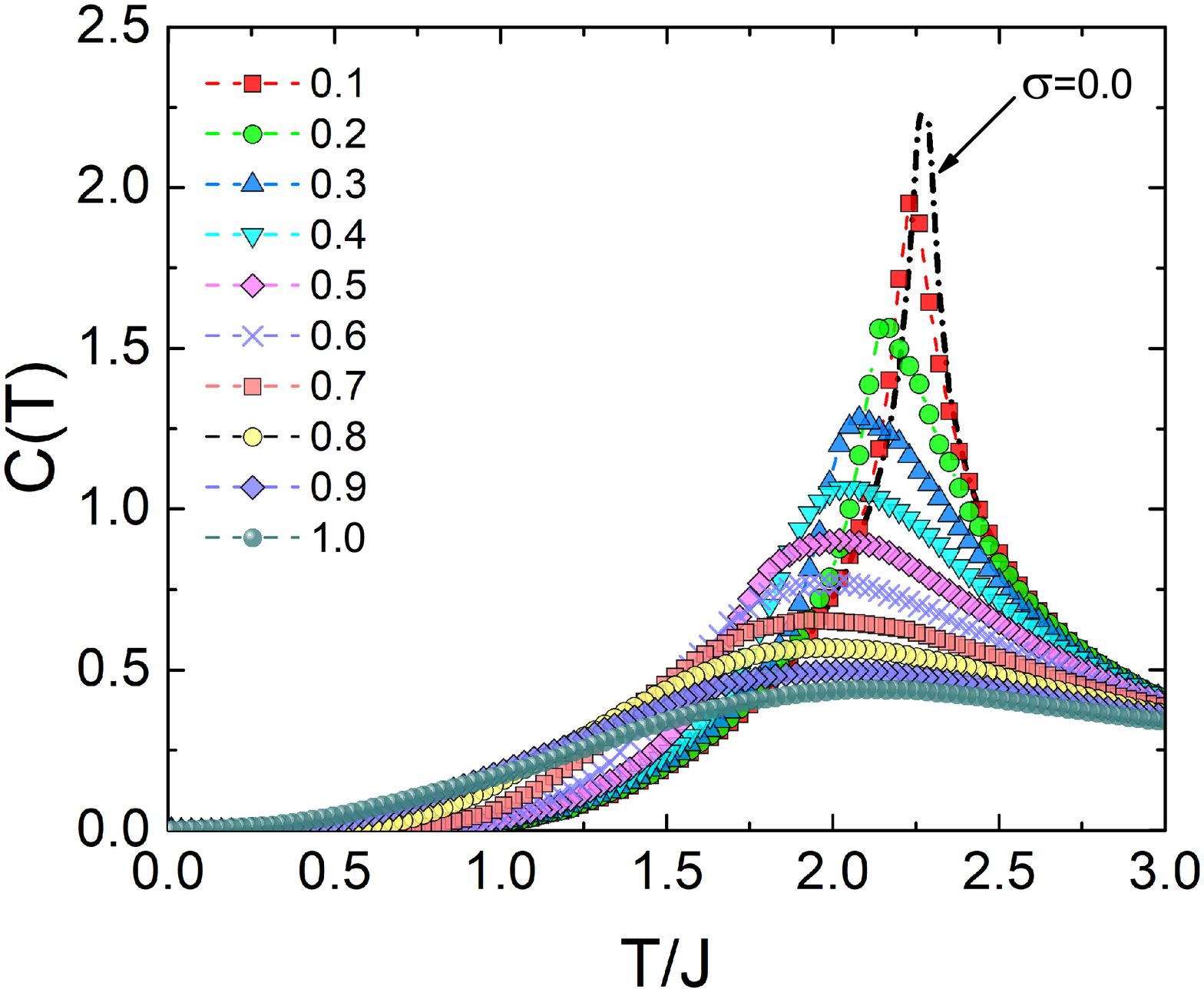}}\\
\caption{(Color online) Thermal dependence of (a) magnetization per spin (b) magnetic specific heat of a square lattice with linear lattice size $L=128$ as 
a function of white noise distribution width $\sigma$ in the absence of external magnetic field. The dashed dotted data correspond to $\sigma=0.0$ case} \label{fig1}
\end{figure}
Let us start our discussion by investigating the effect of noisy magnetic field on the thermal and magnetic properties of the equilibrium version of the present system $(h_{0}=0.0)$. In Fig. \ref{fig1}, we plot the temperature dependence
of magnetization $M(T)$ and specific heat $C(T)$ for a square lattice with $L=128$, corresponding to a variety of values of the randomness parameter $\sigma$. Here, $\sigma=0.0$ case (the dashed curve) 
corresponds to equilibrium 2D Ising model where the magnetization
reduces to zero in the vicinity of the transition temperature $T_{c}$, and the magnetic specific heat exhibits a sharp cusp. It is well known that this model undergoes a continuous phase transition \cite{stanley}. However, as the randomness takes
place with increasing $\sigma$, the magnetization decreases from its saturation value to zero at gradually lower temperature region. Indeed, for moderate $\sigma$ values, it is clear from Fig.\ref{fig1} that the system does not exhibit
critical behavior since the $C(T)$ peak becomes rounded for large $\sigma$.
Hence, we can conclude from this figure that the equilibrium Ising model in the presence of additive white noise does not exhibit conventional order-disorder transitions with increasing randomness.

\begin{figure}[!h]
\center
\subfigure[\hspace{0cm}] {\includegraphics[width=7.0cm]{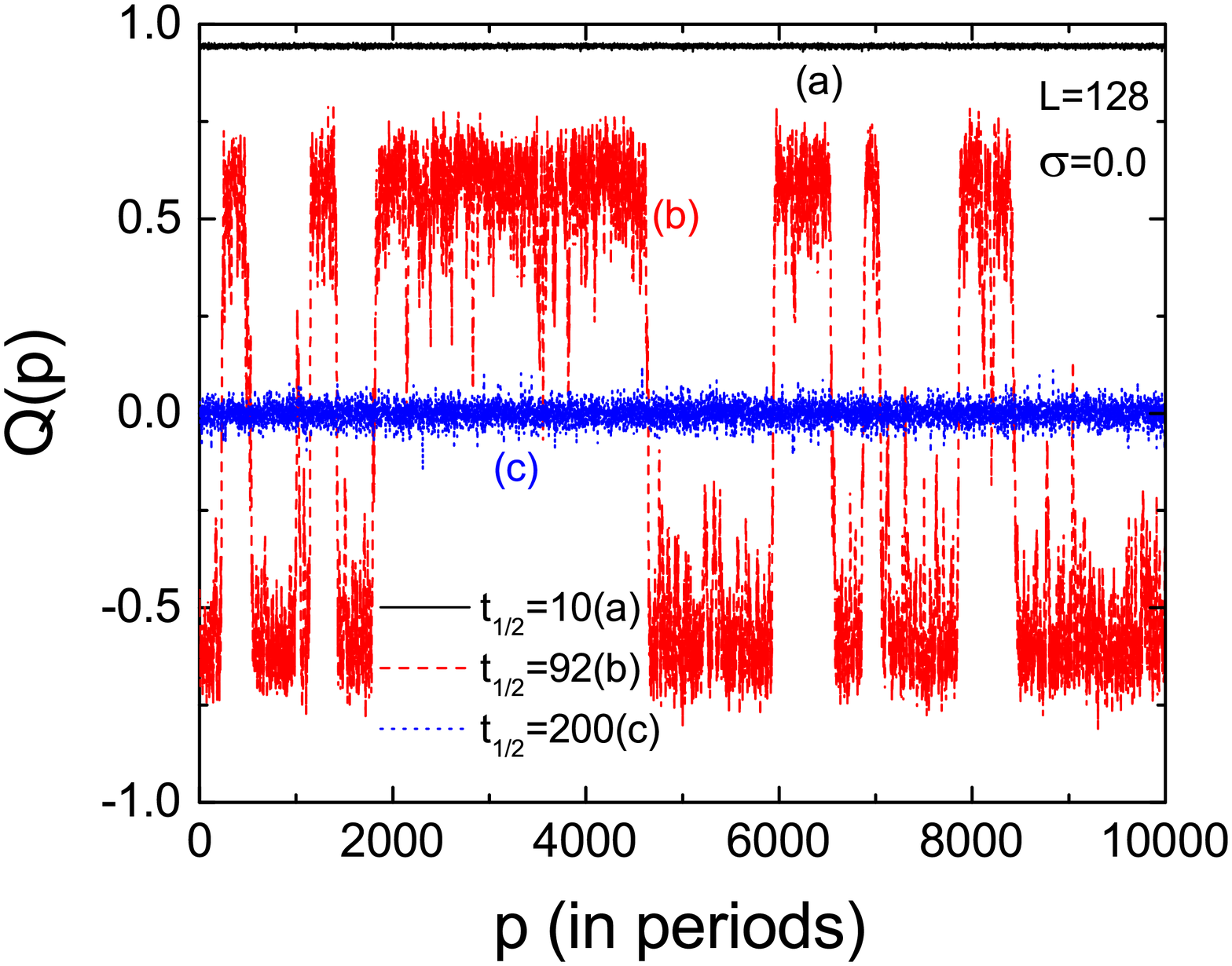}}
\subfigure[\hspace{0cm}] {\includegraphics[width=6.3cm]{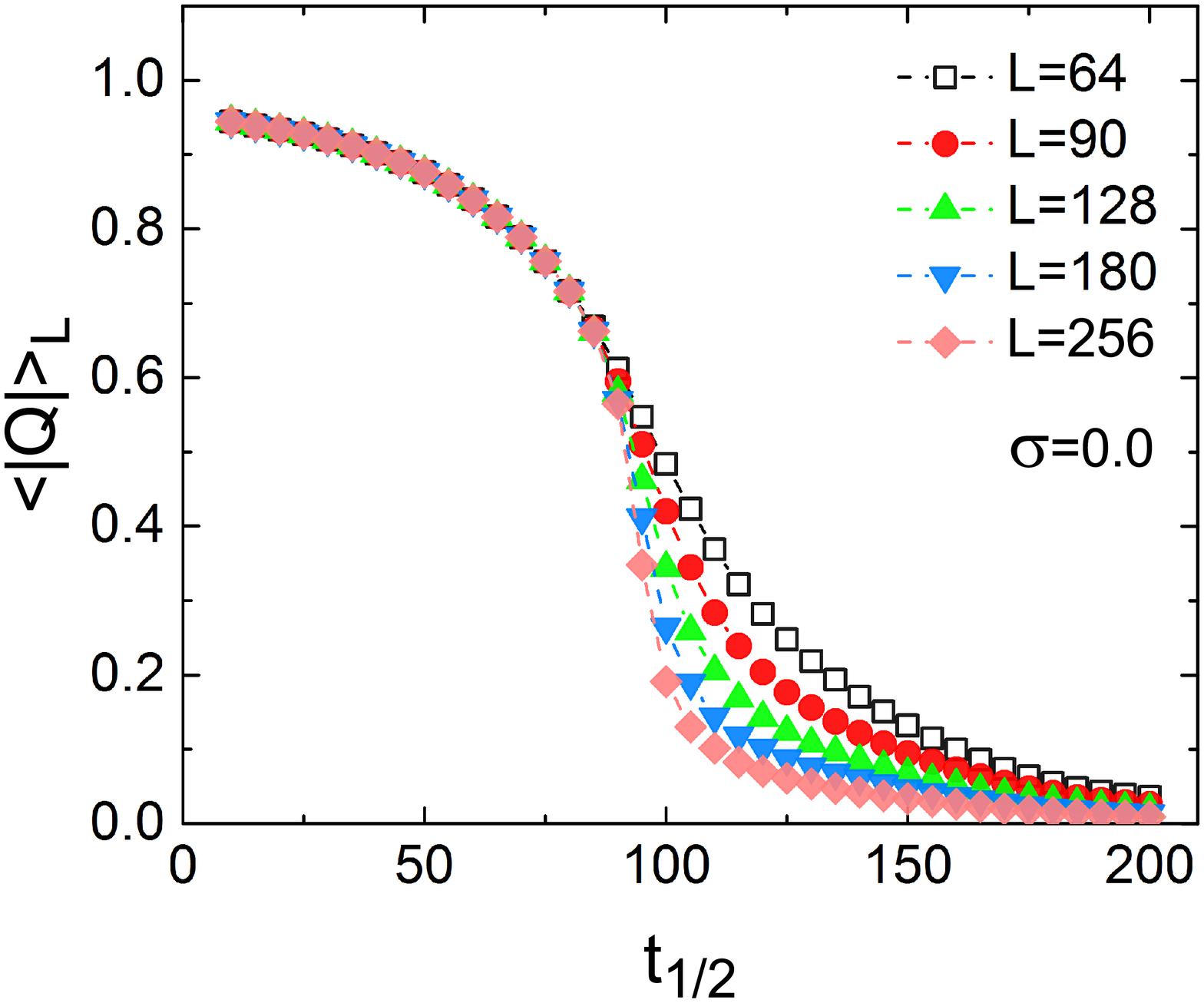}}
\subfigure[\hspace{0cm}] {\includegraphics[width=6.5cm]{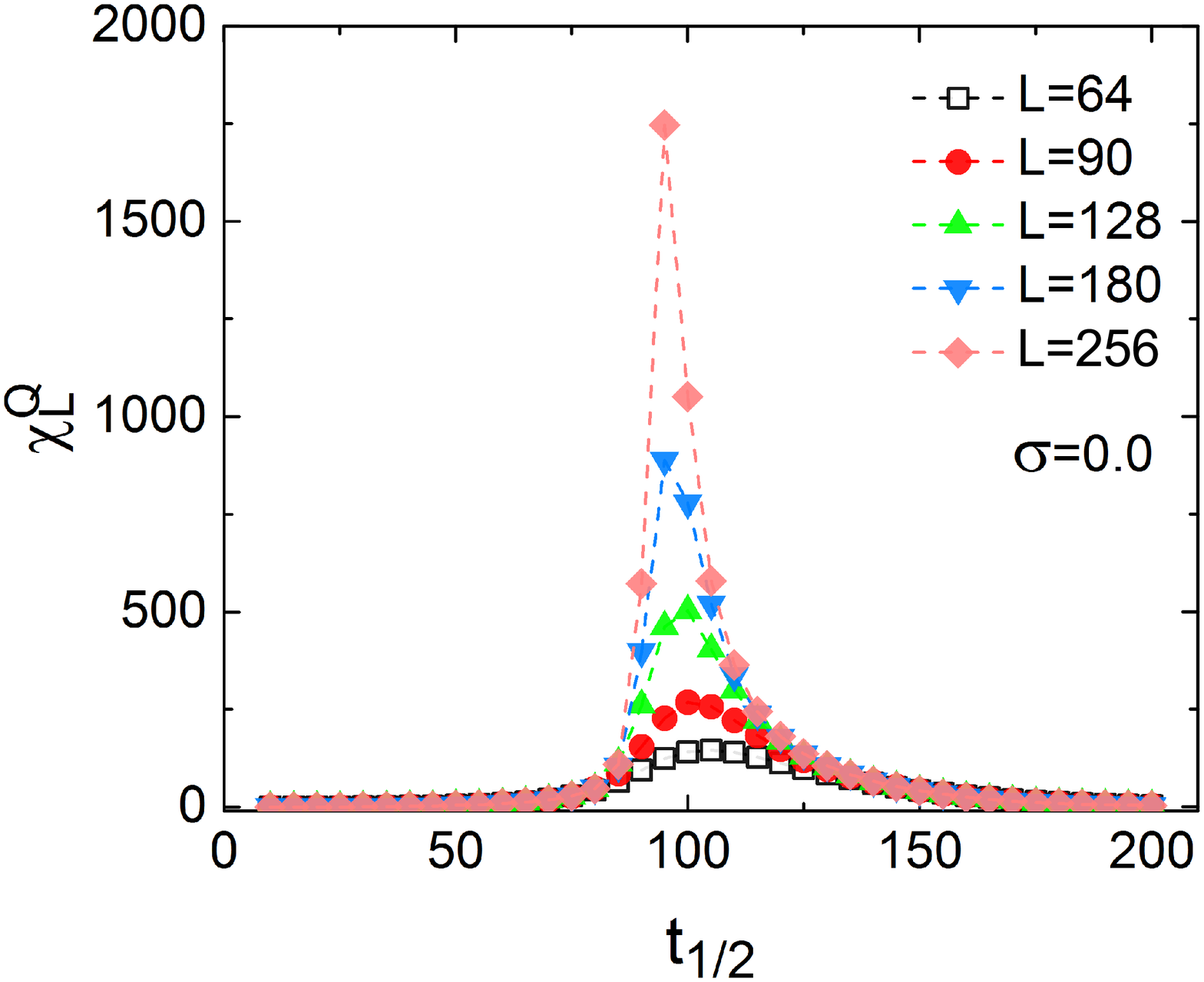}}
\caption{(Color online) (a) Time series of the order parameter $Q$ for $L=128$ and $\sigma=0.0$. In (b) and (c), we present $\langle |Q|\rangle_{L}$ and $\chi^{Q}_{L}$ as a function of half period $t_{1/2}$ for $\sigma=0.0$ for various system sizes.
Each figure has been plotted for $h_{0}=0.3J$ and $T=0.8T_{c}$.} \label{fig2}
\end{figure}
Next, we discuss the DPT properties in the presence of additive white noise. For this aim, defining the half-period parameter $t_{1/2}=0.5P$ is a convenient selection in the literature. 
To out knowledge, DPT properties of kinetic Ising system in the presence of sinusoidally oscillating magnetic field have not been investigated using Metropolis scheme. Hence, 
in Fig. \ref{fig2}, we present our results for a square lattice in the absence of white noise. In Fig. \ref{fig2}a, we present the dynamic order parameter defined by Eq. (\ref{eq7}) as a function of several successive cycles 
of the external field. The behavior of the order parameter $Q$ as a function of magnetic field period is qualitatively the same as that observed in the time series of magnetization $m(t)$ plotted at different temperature values 
for equilibrium system. From Fig. \ref{fig2}a, we see that for high frequency $(t_{1/2}=10)$, the magnetization weakly oscillates around the saturation magnetization whereas for low frequency $(t_{1/2}=200)$, it fluctuates around zero.
For a critical frequency value, large fluctuations with highly stochastic nature originate. From this, one can see that the system exhibits a frequency (or period) induced dynamic phase transition. In order to obtain 
further information about this DPT behavior, we present the dependence of the order parameter $\langle |Q|\rangle_{L}$, and the scaled variance $\chi_{L}^{Q}$ as a function of the half period $t_{1/2}$ in Fig. \ref{fig2}b and \ref{fig2}c, respectively.
By keeping in mind the pronounced finite size effects, it is apparent that $\langle |Q|\rangle_{L}$ evolves from saturation value to zero with increasing $t_{1/2}$. In addition to this, $\chi_{L}^{Q}$ plotted for large system sizes such as 
$L=256$ exhibits a prominent peak around the critical value $t_{1/2}^{c}$. 

\begin{figure}[!h]
\center
\subfigure[\hspace{0cm}] {\includegraphics[width=6.5cm]{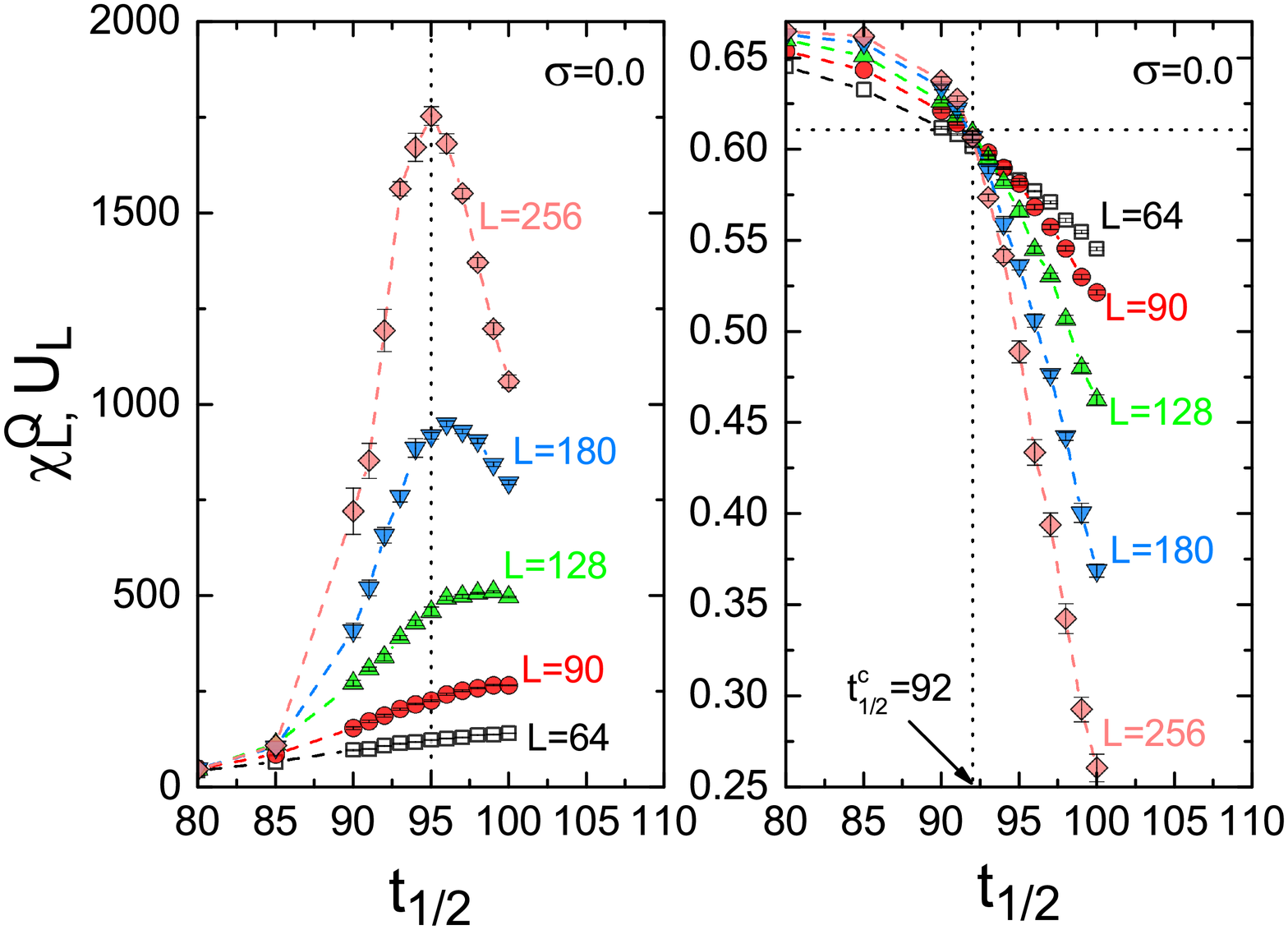}}
\subfigure[\hspace{0cm}] {\includegraphics[width=6.5cm]{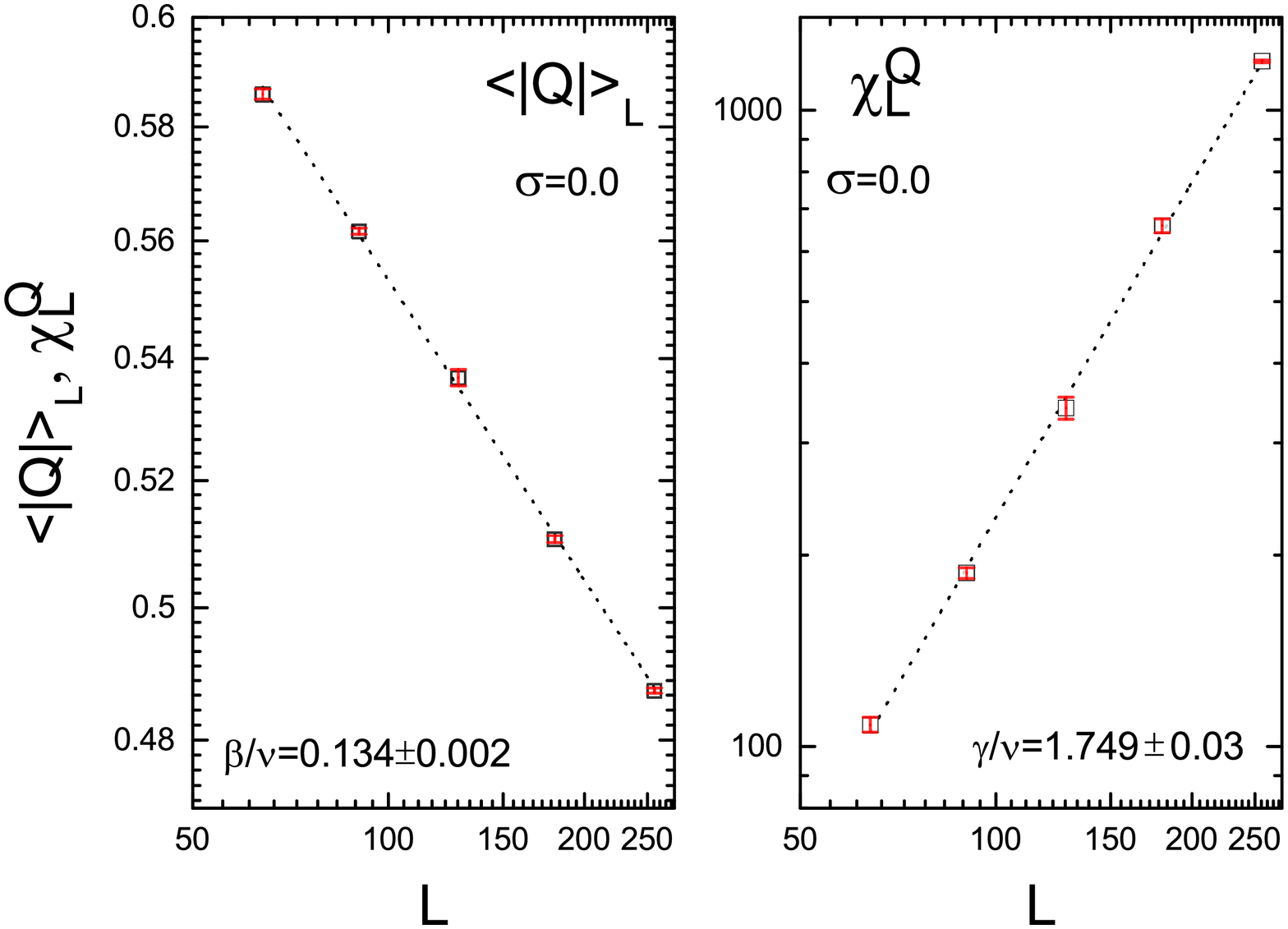}}\\
\caption{(Color online) (a) Half period $t_{1/2}$ dependence of scaled variance $\chi^{Q}_{L}$ and Binder cumulant $U_{L}$ for $\sigma=0.0$. $\chi^{Q}_{L}$ is maximized at $t_{1/2}=95$. Binder cumulant curves intersect each other at $t_{1/2}^{c}=92$. 
The horizontal dotted line corresponds to $U_{L}^{c}=0.61069$ \cite{kamieniarz} for Ising model. (b) Log-log plots regarding the estimated critical exponents of dynamic order parameter 
$\langle |Q|\rangle_{L}$ and scaled variance $\chi^{Q}_{L}$ as functions of $L$ at $t_{1/2}=92$. The dotted lines denote the fitting result. Note that the error bars are smaller than the data symbols.} \label{fig3}
\end{figure}
\begin{table}[h]
\begin{center}
\begin{threeparttable}
\caption{Comparison of critical properties of kinetic Ising model obtained by a variety of works and those obtained in the present work. The abbreviated algorithm names 
respectively denote soft Glauber (SGL.), Glauber (GL.), and Metropolis (MPL.)}
\renewcommand{\arraystretch}{1.3}
\begin{tabular}{ccccccccccccc}
\thickhline
Reference &\cite{robb}       &\cite{sides3} & \cite{buendia} &\cite{vatansever} & \cite{robb2} & \cite{korniss}&\cite{sides5}& \cite{park} & Present Work\\
\hline 
$h(t)$ &sawtooth & sine &  square & square & square & square &sine & square & sine \\
lattice  $(q)$  &4 & 4 & 4 & 6 (tria.)  & 4 & 4 & 4 & 6 (SC) & 4 \\
algorithm & GL & GL. & SGL. & MPL.  & GL.  & GL. & GL. & - & MPL. \\
$P_{c}$ (MCSS)& 493 & 256 &  290 & 284 & 137 & 140 &-& 121 & 184 \\
$\beta/\nu$ &- & 0.11 &  0.144 & 0.143 & 0.126 & 0.126 &0.11 & 0.51 & 0.134 \\
$\gamma/\nu$ $(t_{1/2}^{c})$& - & 1.84 &  1.77 & 1.77 & - & 1.74 &-& - & - \\
$\gamma/\nu$ $(\chi^{peak})$& - & 1.84 &  1.79 & 1.75 &1.74  & 1.78 &1.84 & 1.96 & 1.75 \\
\thickhline \\
\end{tabular}\label{table1}
\end{threeparttable}
\end{center}
\end{table}
In order to precisely measure the critical value $t_{1/2}^{c}$, we have calculated the Binder cumulant $U_{L}$ within a narrow region of the half period such as $80\leq t_{1/2}\leq100$. The results are displayed in Fig. \ref{fig3}a from which 
we observe a perfect intersection of $U_{L}$ curves corresponding to different values of linear lattice size $L$. Our numerical data yield $U_{L}^{*}=0.61$ and $t_{1/2}^{c}=92$. 
Note that the value $t_{1/2}=95$ at which $\chi^{Q}_{L}$ diverges does not coincide with the value obtained from Binder cumulant analysis. We also underline that our estimated value $U_{L}^{*}=0.61$
for the kinetic Ising model can be compared to that obtained for the equilibrium model $U_{L}^{c}=0.61069$ \cite{kamieniarz}.  The critical period value $P_{c}=2t_{1/2}=184$ is a new value in the literature. Using the scaling relations
for the dynamic order parameter and scaled variance  \cite{buendia},
\begin{equation}\label{eq11}
\langle |Q|\rangle_{L}\propto L^{-\beta/\nu},  
\end{equation}
\begin{equation}\label{eq12}
\chi_{L}^{Q}\propto L^{\gamma/\nu},  
\end{equation}
we have also extracted the critical exponent ratios $\beta/\nu$ and $\gamma/\nu$ at $P_{c}=2t_{1/2}=184$. The obtained results have been shown in Fig. \ref{fig3}b. Our estimate for the exponent ratios $\beta/\nu=0.134$, and $\gamma/\nu=1.749$ 
are clearly very close to those obtained for the equilibrium counterpart of the 2D Ising model \cite{stanley}. Combining our results in the following scaling \cite{sides4} relation we get
\begin{equation}\label{eq13}
2(\beta/\nu)+\gamma/\nu=2.018\approx d. 
\end{equation}
Based on the above discussions, we can clearly say that DPT observed in the present system in the absence of white noise falls into the same universality class with its equilibrium counterpart. The critical values obtained 
in the present analysis have been presented in the Table \ref{table1} along with the results of the previous works.

\begin{figure}[!h]
\center
\subfigure[\hspace{0cm}] {\includegraphics[width=6.5cm]{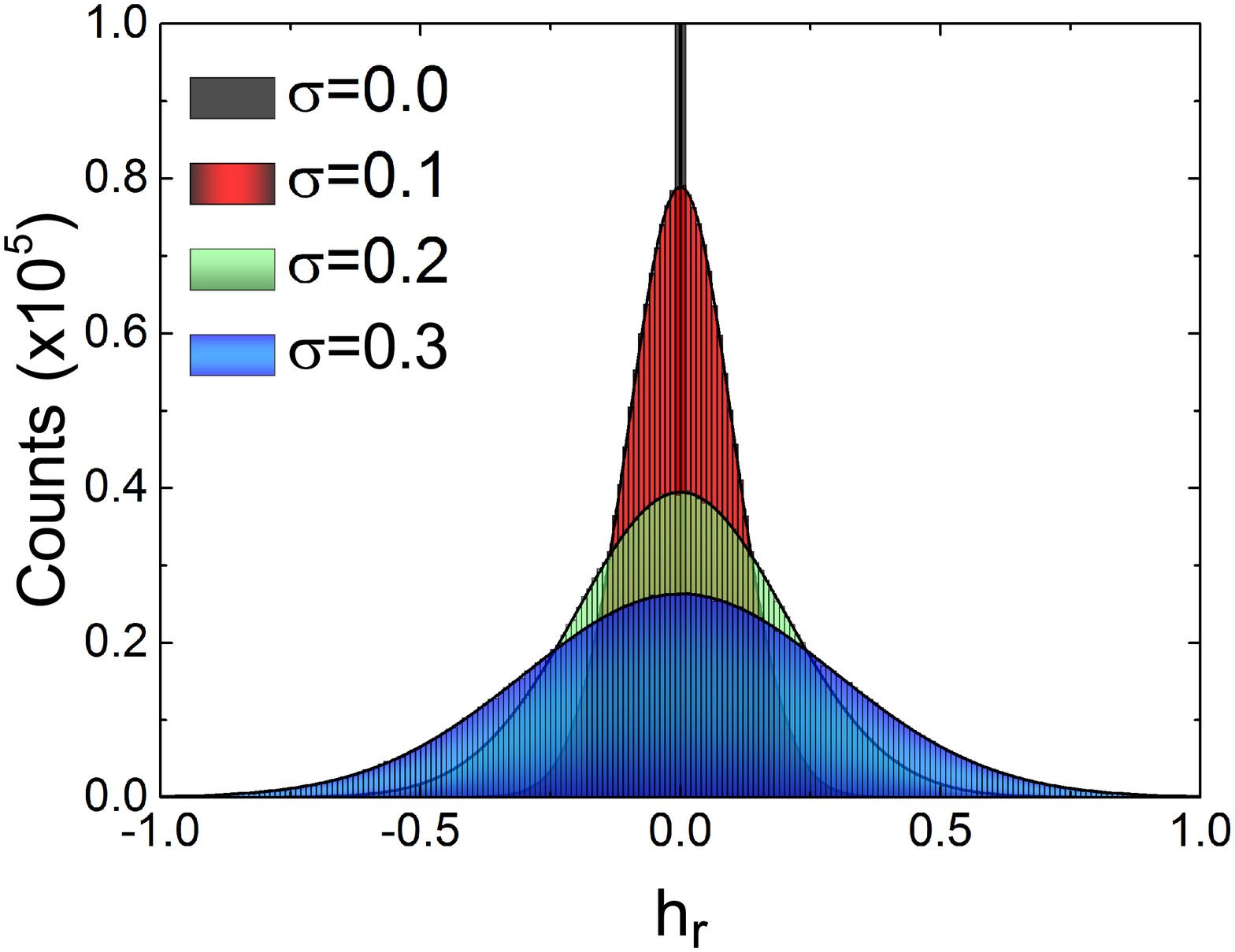}}
\subfigure[\hspace{0cm}] {\includegraphics[width=6.5cm]{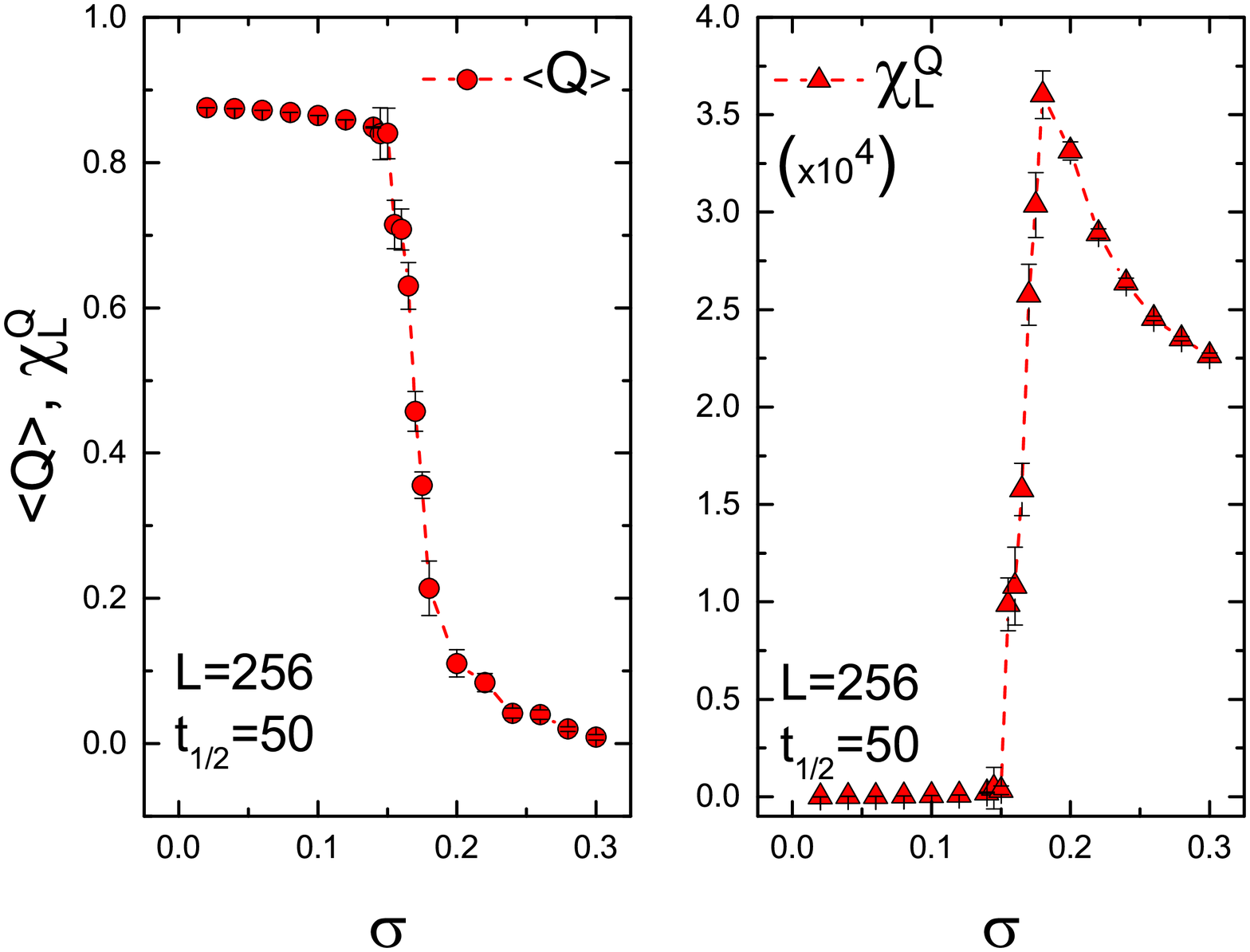}}\\
\subfigure[\hspace{0cm}] {\includegraphics[width=6.5cm]{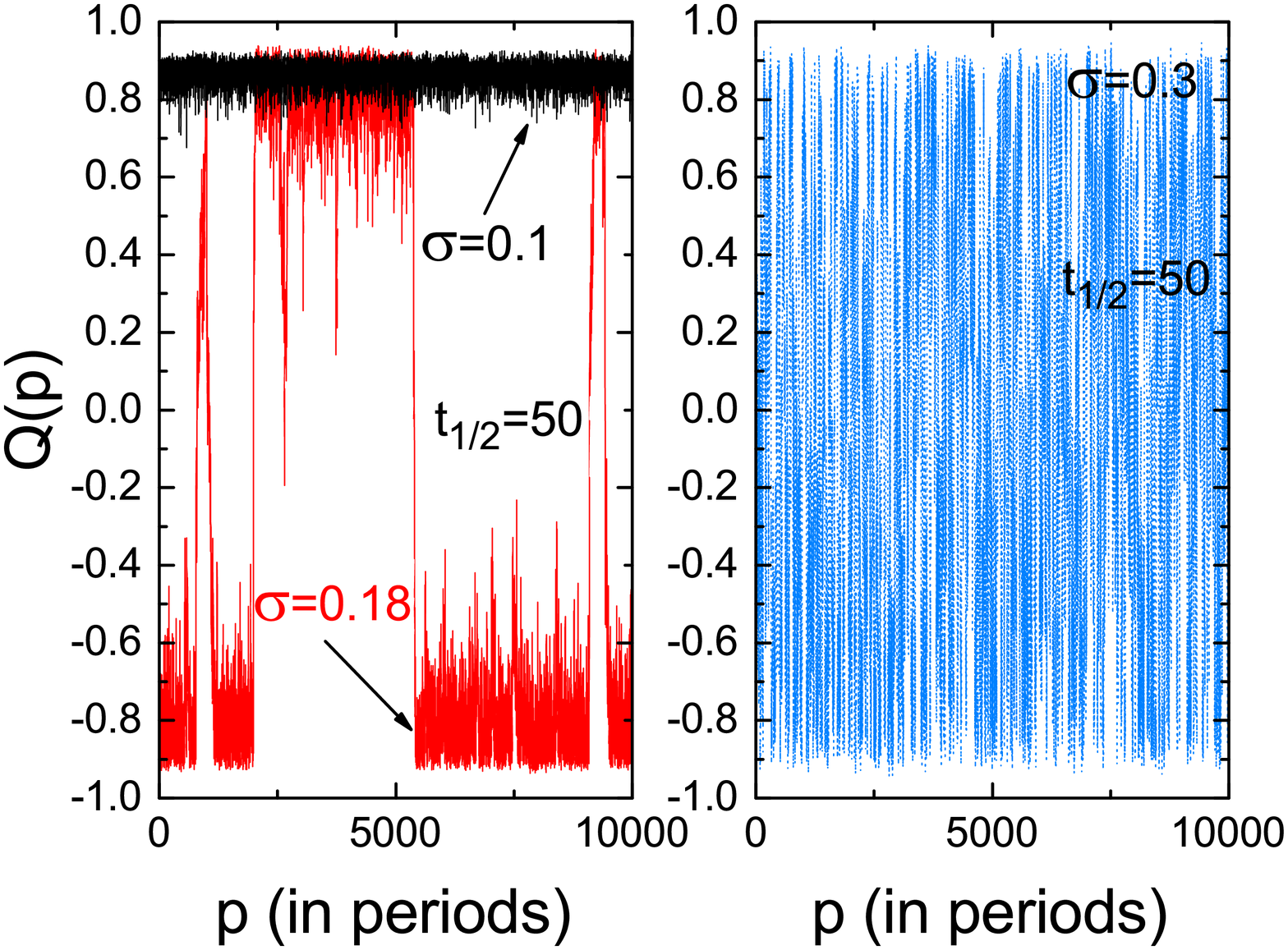}}
\caption{(Color online) (a) Simulated distribution of Gaussian white noise corresponding to various $\sigma$ values. (b) Variation of $\langle Q\rangle$ and $\chi^{Q}_{L}$ as functions of distribution width $\sigma$ for $L=256$ and $t_{1/2}=50$. 
(c) Time series of the order parameter $Q$ for $L=256$ and $t_{1/2}=50$ with $\sigma=0.1$, $0.18$, and $0.3$.} \label{fig4}
\end{figure}
\begin{figure}[!h]
\center
\subfigure[\hspace{0cm}] {\includegraphics[width=4.0cm]{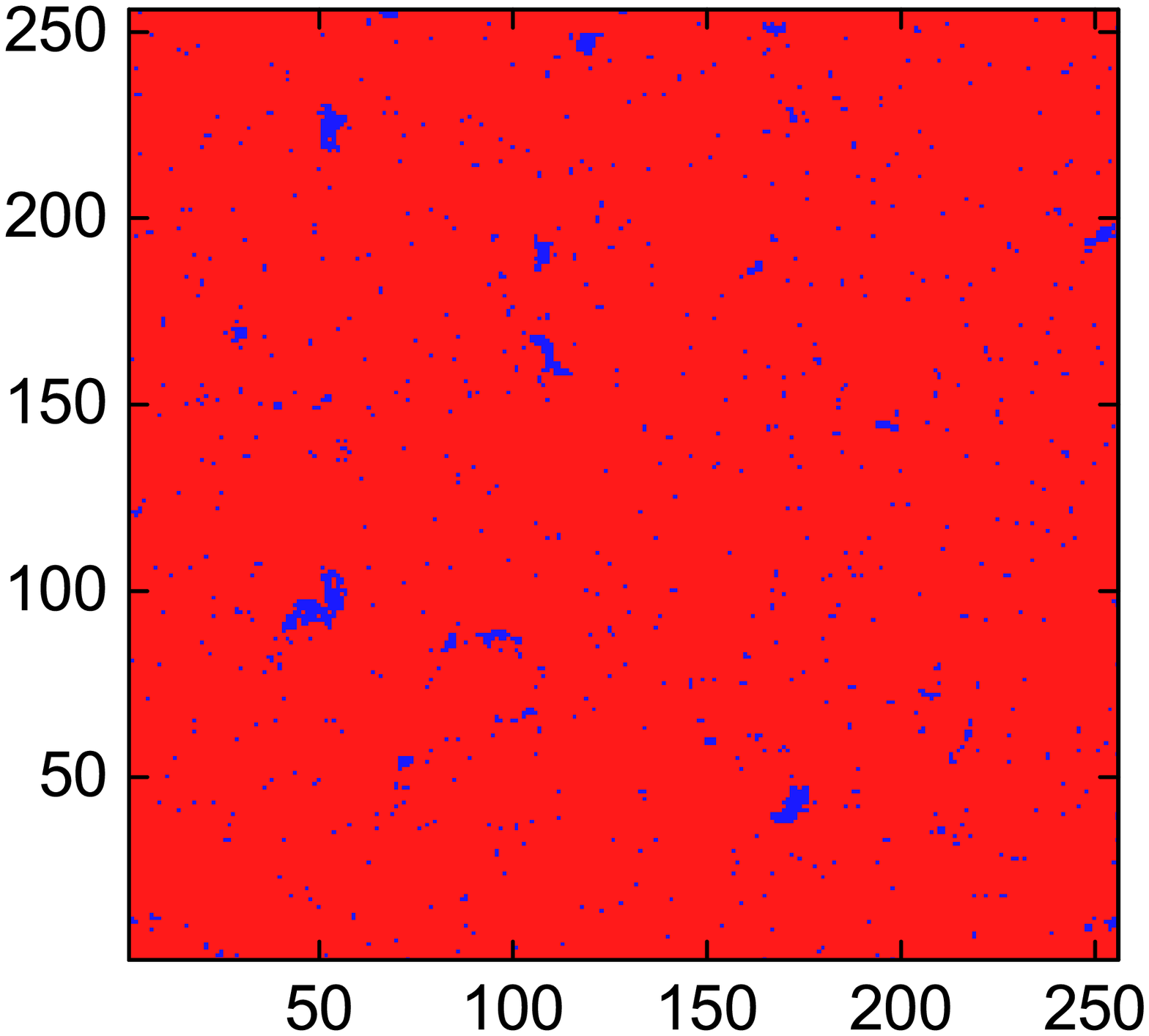}}
\subfigure[\hspace{0cm}] {\includegraphics[width=4.0cm]{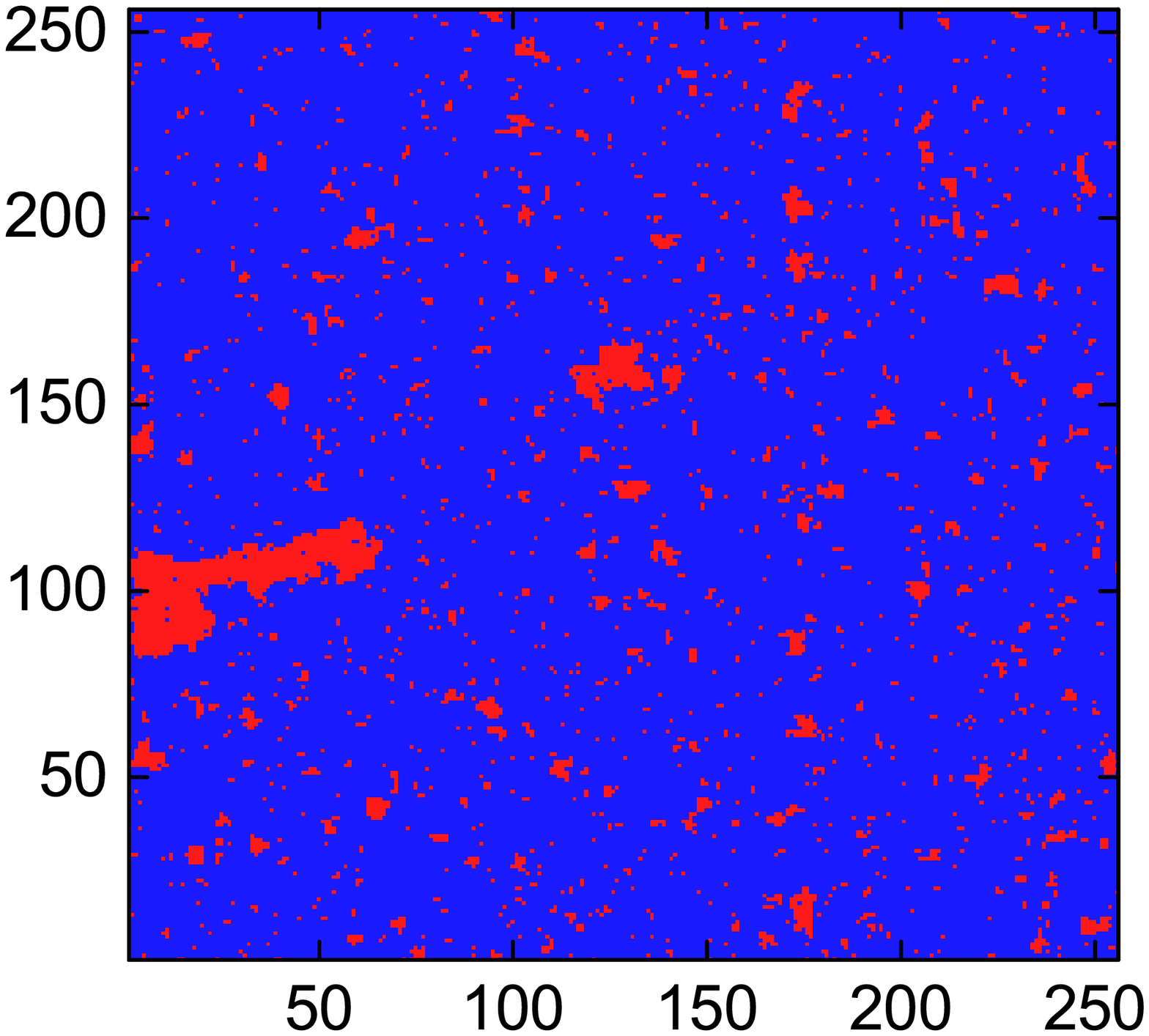}}
\subfigure[\hspace{0cm}] {\includegraphics[width=4.0cm]{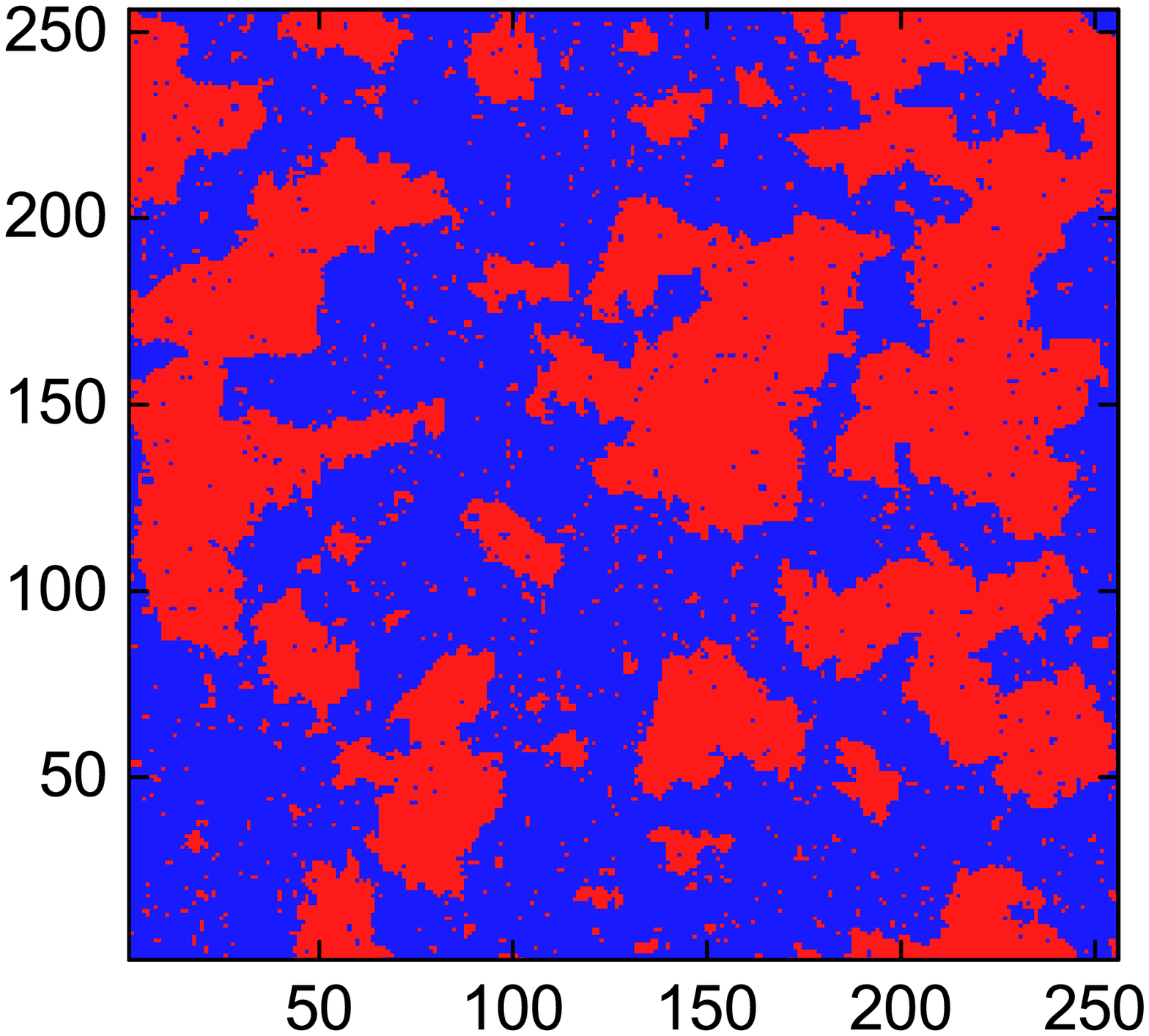}}\\
\caption{(Color online) Snapshots of the kinetic Ising model with $L=256$ and $t_{1/2}=50$ in the presence of additive white noise. The amount of randomness is determined by the 
Gaussian width $\sigma$: (a) $\sigma=0.1$, (b) $\sigma=0.18$, (c) $\sigma=0.3$. Red and blue regions respectively denote spin-$\uparrow$ (+1), and spin-$\downarrow$ (-1) states.} \label{fig5}
\end{figure}
In order to investigate the influence of the additive white noise on the DPT properties of the kinetic Ising model, we have plotted in Fig. \ref{fig4} the variation of dynamic order 
parameter $\langle |Q|\rangle_{L}$, and the scaled variance $\chi^{Q}_{L}$
as functions of the noise parameter $\sigma$ for a square lattice with $L=256$, and half period value $t_{1/2}=50$. The simulated random field distributions within $0.0\leq \sigma\leq0.3$ are also depicted in this figure. One can clearly
deduce from Fig. \ref{fig4}b that the order parameter $\langle |Q|\rangle_{L}$ evolves from unity to zero with increasing randomness. Besides, the variance $\chi^{Q}_{L}$ passes through  a sharp maximum at a critical randomness $\sigma_{c}$. 
Evolution of $Q$ as a function of field period $P$ also supports the possible presence of disorder induced phase transition in the system. However, there is a distinction which shows itself for large randomness such as $\sigma=0.3$ where 
$Q(p)$ randomly oscillates between states $\pm1$. The snapshots of the system corresponding to $\sigma$ values given in Fig. \ref{fig4}c have been presented in Fig. \ref{fig5}. Fig. \ref{fig5}a shows that in the presence of weak randomness 
such as $\sigma=0.1$, the system exhibits a long range order where the great majority of the spins pointing in the spin-$\uparrow$ direction. Form Fig. \ref{fig5}b, 
we see that large domains of the same oriented spins are formed for moderate values of $\sigma$. According Fig. \ref{fig5}c, although the dynamic order parameter tends to zero for large randomness, instead of nucleated droplets, we observe
large domains of spins, hence a short range order originates in the system. Due to these large fluctuations and strong stochastic behavior for large randomness, we will restrict our subsequent investigations to weak noise case.  

\begin{figure}[!h]
\center
\subfigure[\hspace{0cm}] {\includegraphics[width=6.5cm]{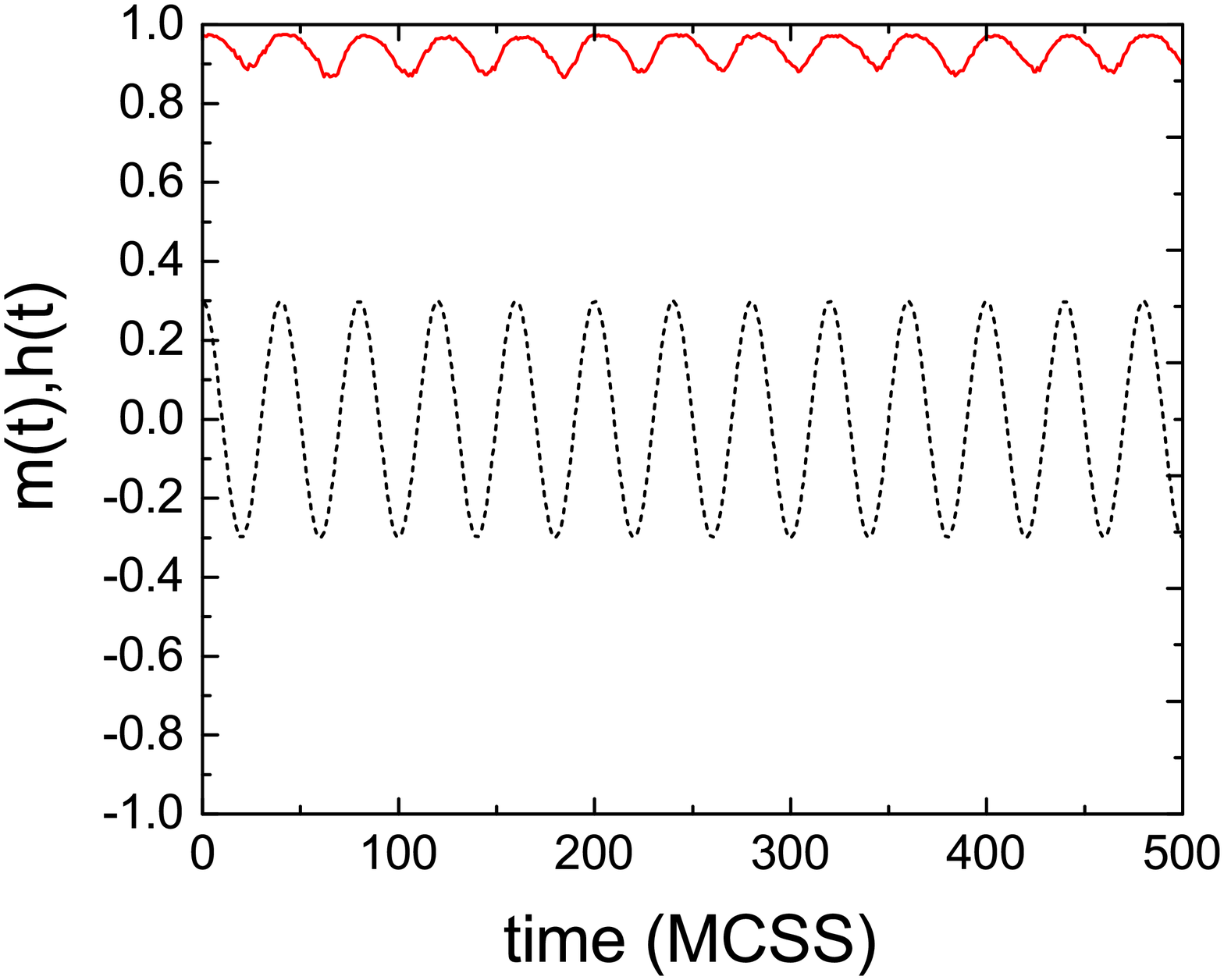}}
\subfigure[\hspace{0cm}] {\includegraphics[width=6.5cm]{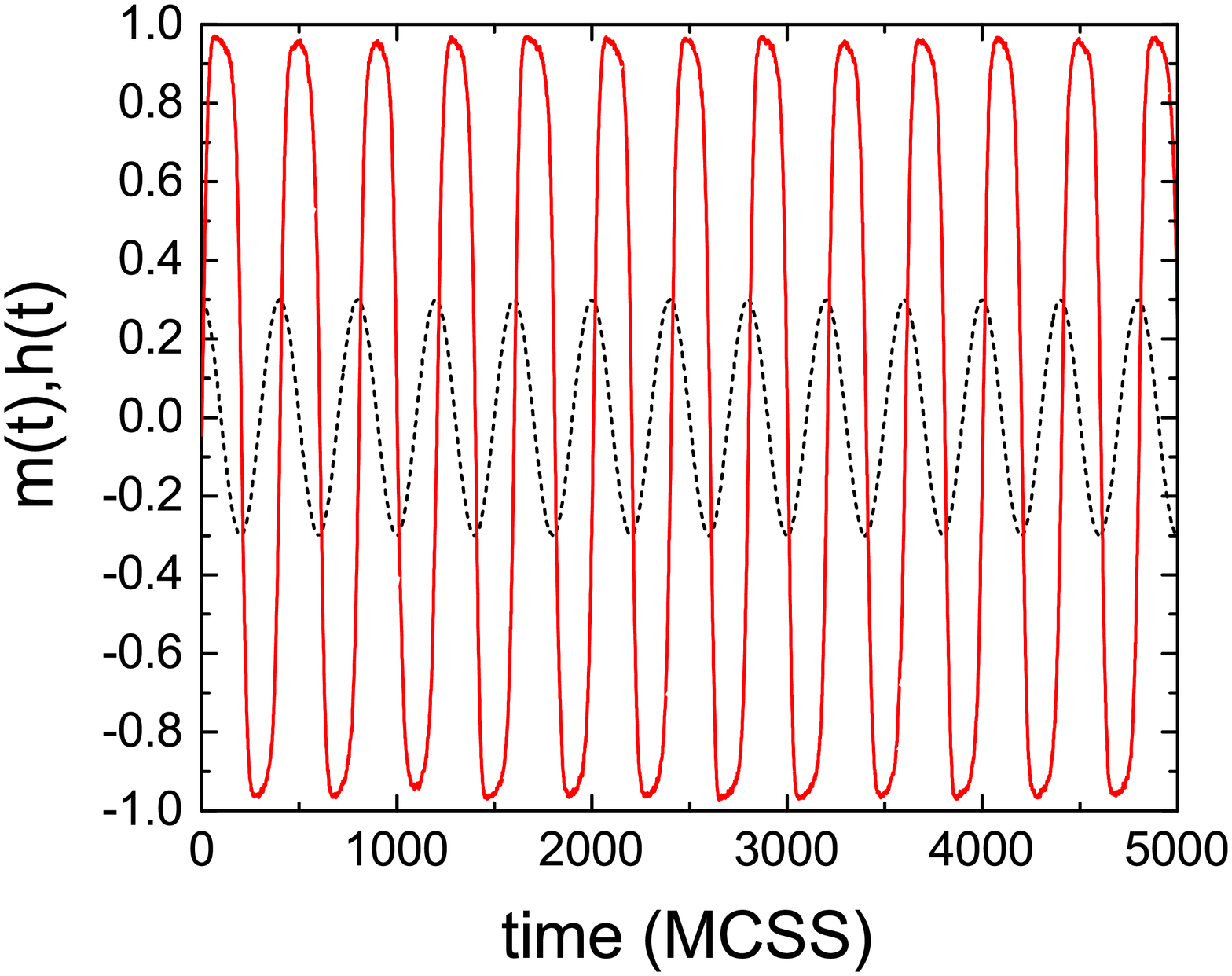}}\\
\subfigure[\hspace{0cm}] {\includegraphics[width=6.5cm]{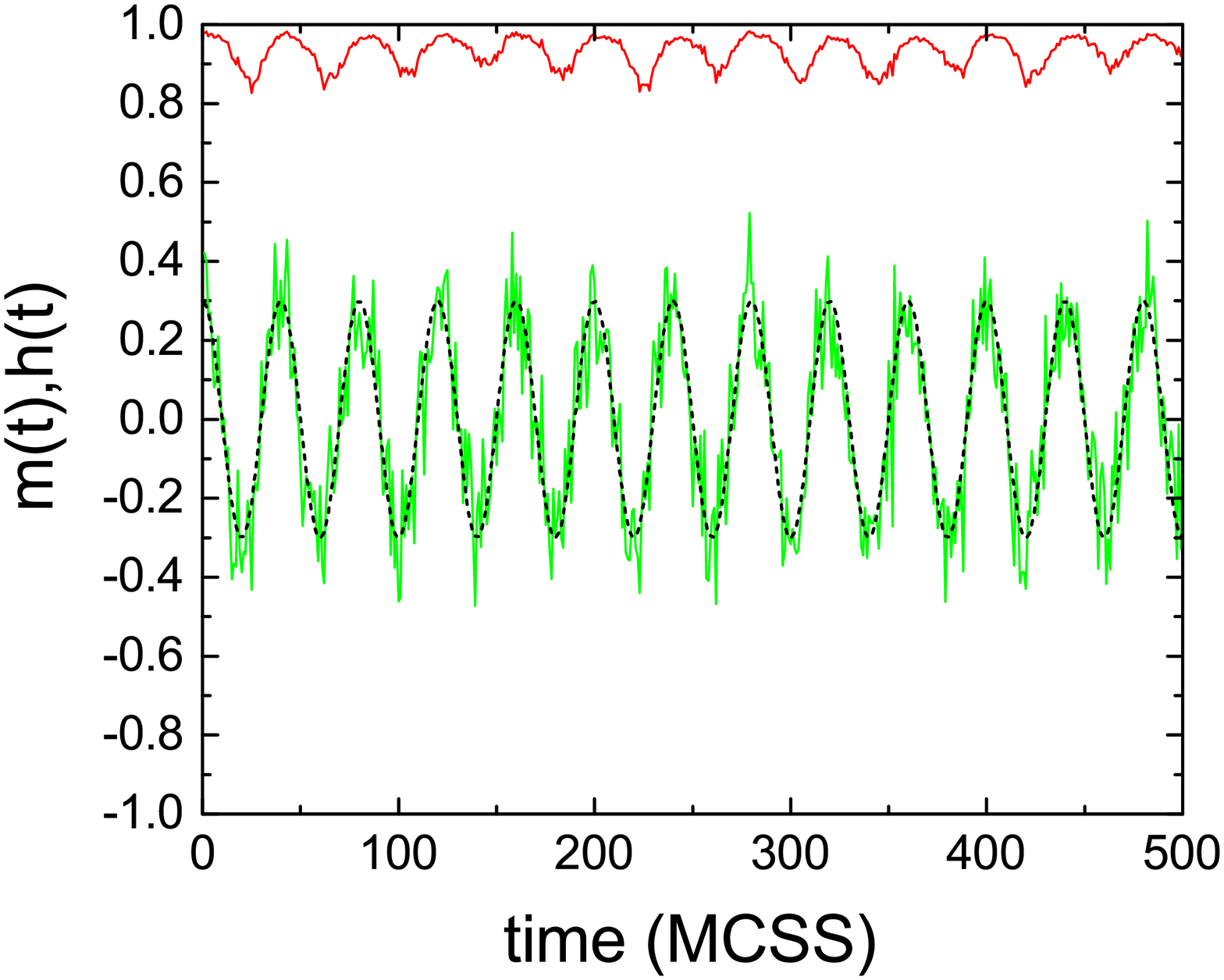}}
\subfigure[\hspace{0cm}] {\includegraphics[width=6.5cm]{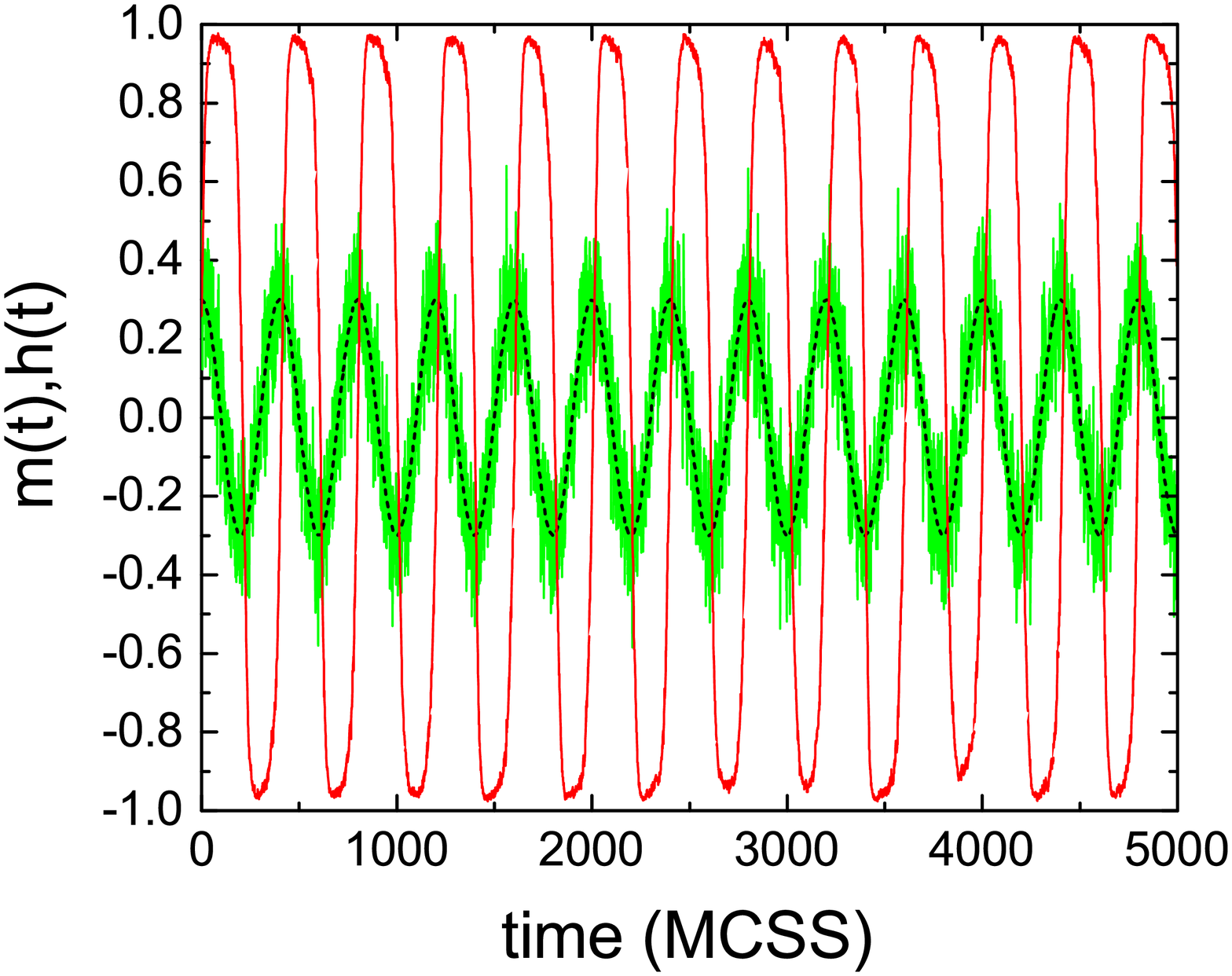}}\\
\caption{(Color online) Time series of instantaneous magnetization $m(t)$, and magnetic field $h(t)$ for $h_{0}=0.3J$ and $L=128$ where the time is defined in units of MCSS. Simulation parameters are as follows: 
(a) $t_{1/2}=20$, $\sigma=0.0$; (b) $t_{1/2}=200$, $\sigma=0.0$; (c) $t_{1/2}=20$, $\sigma=0.1$; (a) $t_{1/2}=200$, $\sigma=0.1$. Green signal is the noisy magnetic field with $\sigma=0.1$.} \label{fig6}
\end{figure}
\begin{figure}[!h]
\center
\subfigure[\hspace{0cm}] {\includegraphics[width=7.0cm]{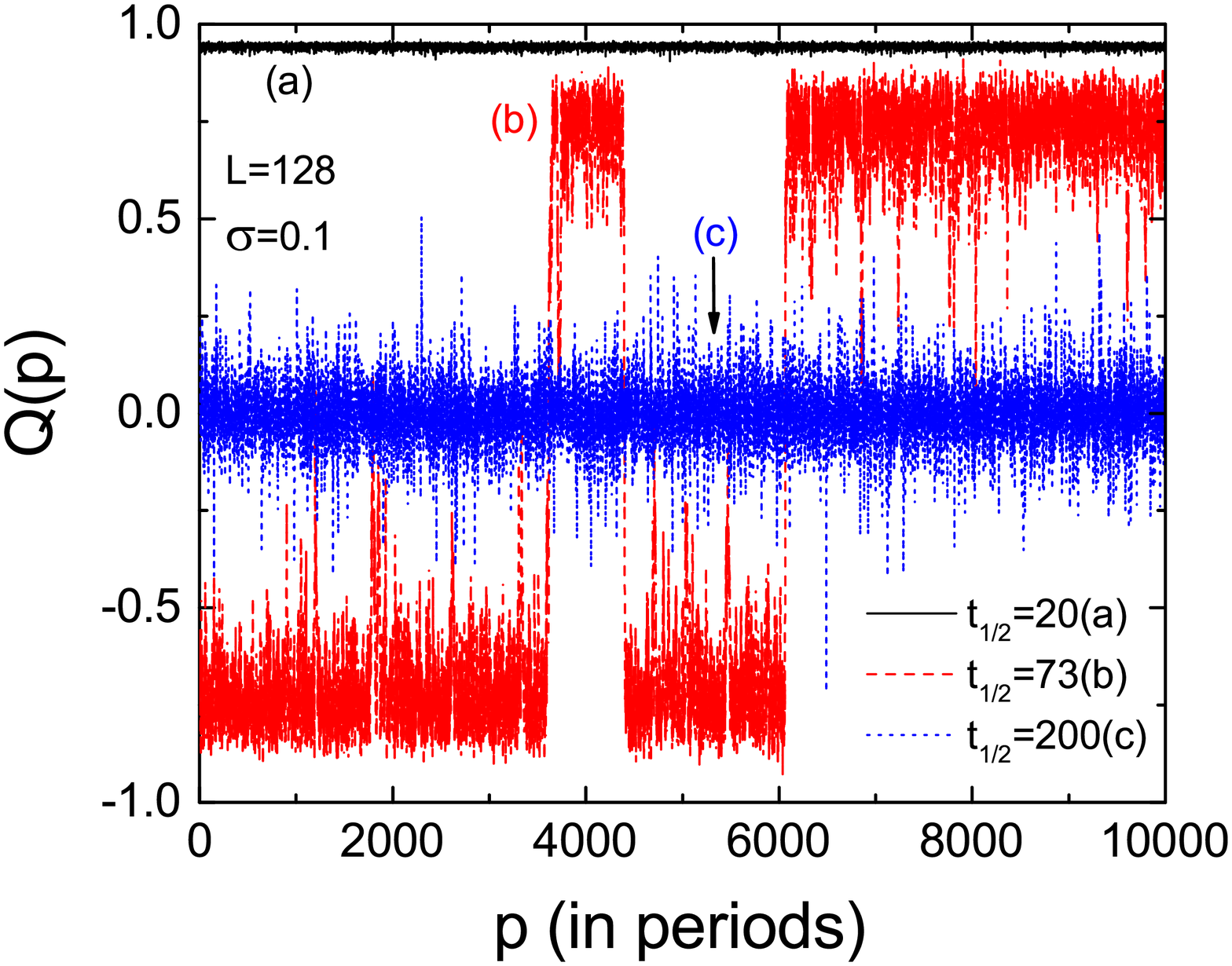}}
\subfigure[\hspace{0cm}] {\includegraphics[width=6.3cm]{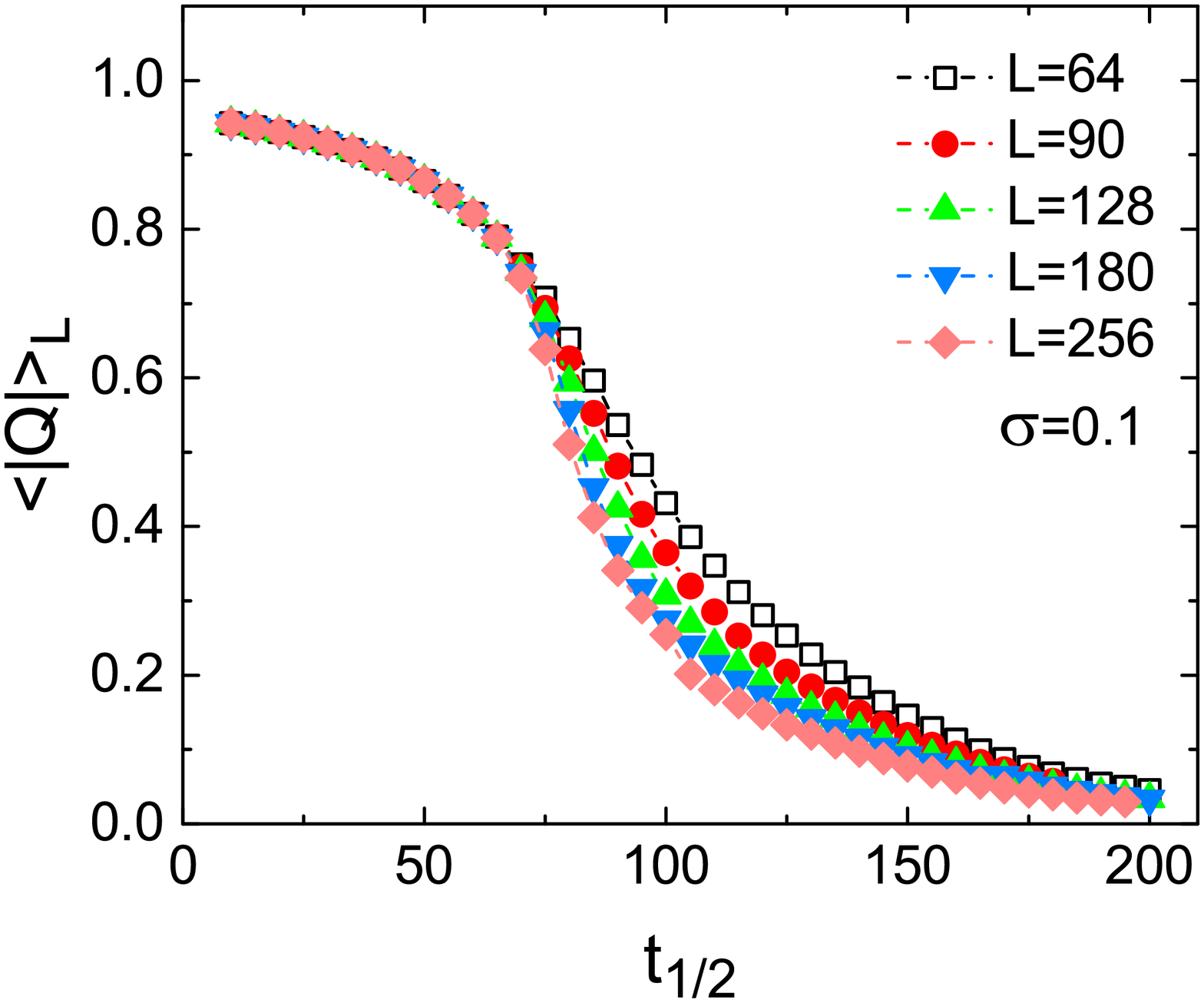}}
\subfigure[\hspace{0cm}] {\includegraphics[width=6.5cm]{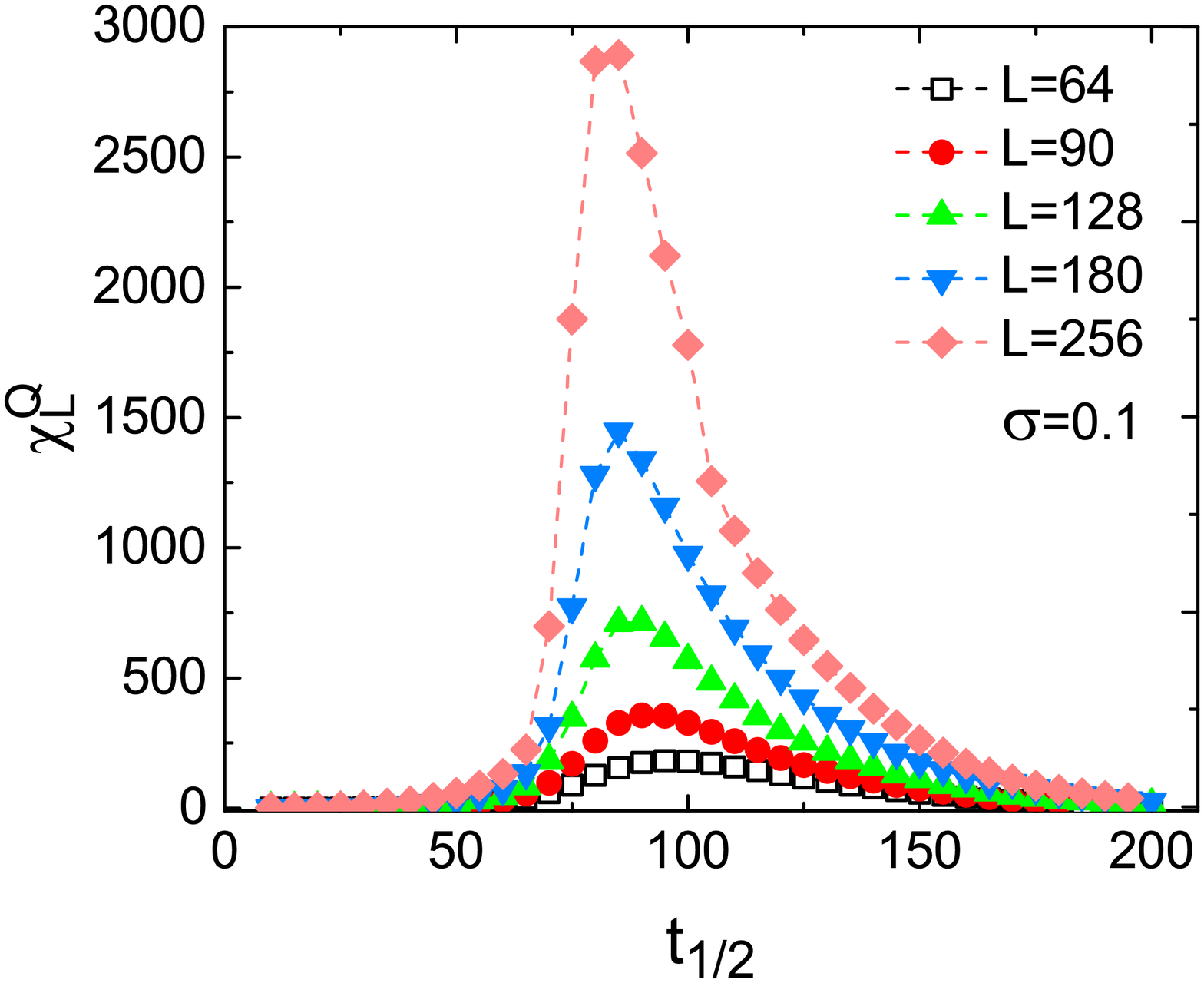}}
\caption{(Color online) (a) Time series of the order parameter $Q$ for $L=128$ and $\sigma=0.1$. In (b) and (c), we present $\langle |Q|\rangle$ and $\chi^{Q}_{L}$ as a function of half period $t_{1/2}$ for $\sigma=0.1$ for various system sizes.
Each figure has been plotted for $h_{0}=0.3J$ and $T=0.8T_{c}$.} \label{fig7}
\end{figure}
\begin{figure}[!h]
\center
\subfigure[\hspace{0cm}] {\includegraphics[width=6.5cm]{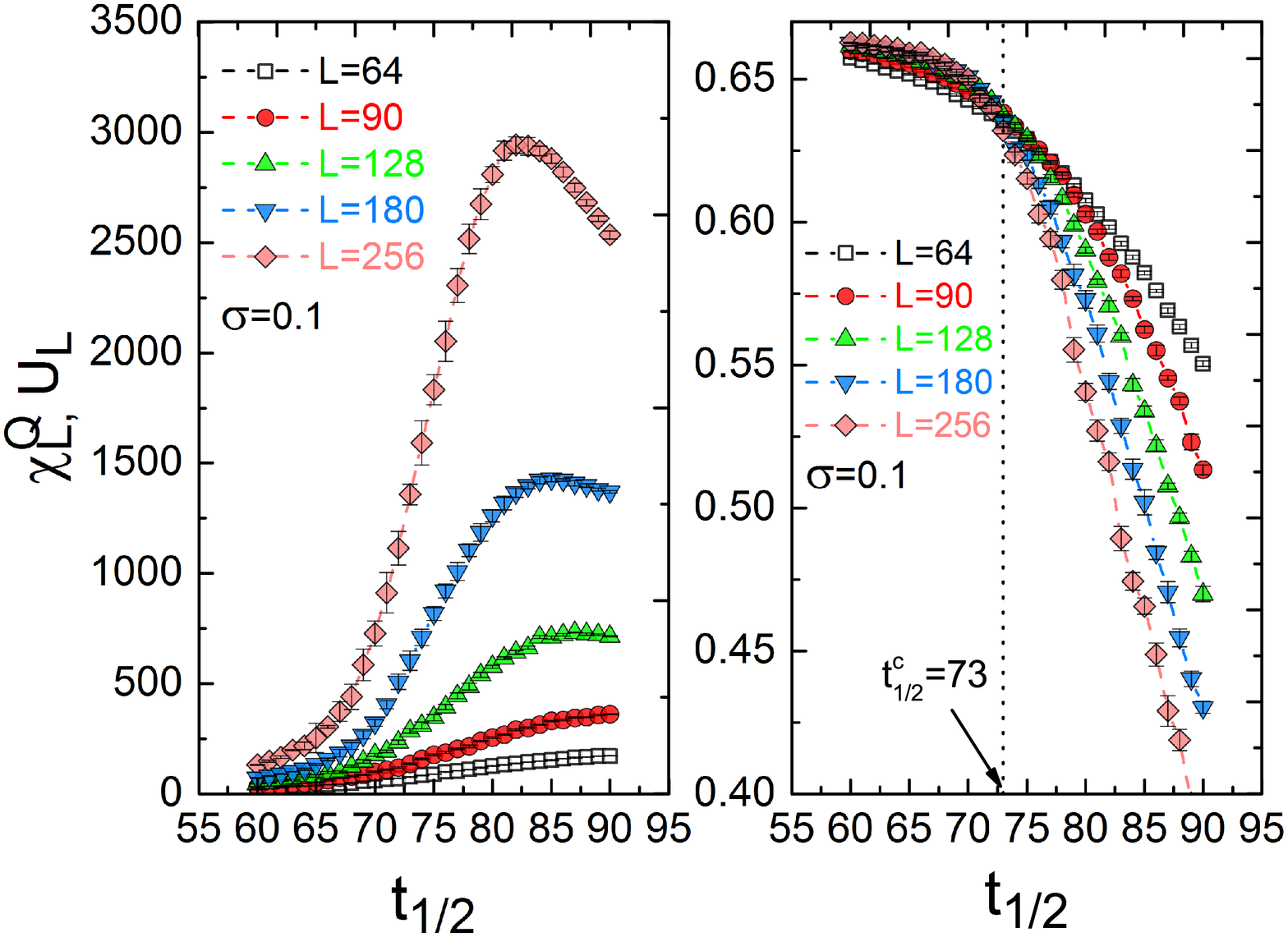}}
\subfigure[\hspace{0cm}] {\includegraphics[width=6.5cm]{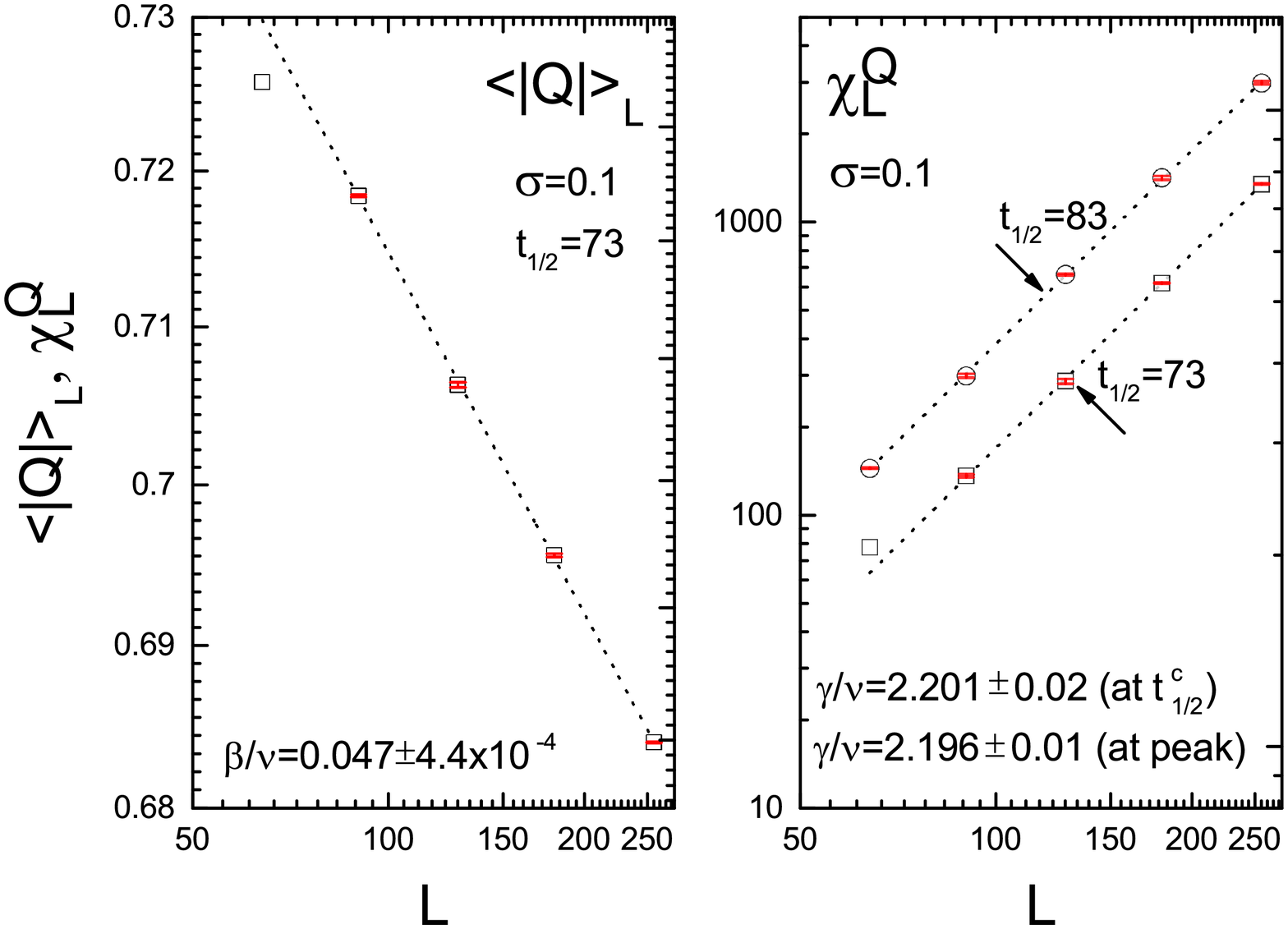}}
\caption{(Color online) (a) Half period $t_{1/2}$ dependence of scaled variance $\chi^{Q}_{L}$ and Binder cumulant $U_{L}$ for $\sigma=0.1$. $\chi^{Q}_{L}$ is maximized at $t_{1/2}=83$. Binder cumulant curves intersect each other at $t_{1/2}^{c}=73$. 
The horizontal dotted line corresponds to $U_{L}^{c}=0.61069$ \cite{kamieniarz} for Ising model with $\sigma=0.0$. (b) Log-log plots regarding the estimated critical exponents of dynamic order parameter 
$\langle |Q|\rangle$ and scaled variance $\chi^{Q}_{L}$ as functions of $L$ at $t_{1/2}=83$ (peak of the variance) and $t_{1/2}=73$ (Binder cumulant intersection). 
The dotted lines denote the fitting result. Note that the error bars are smaller than the data symbols.} \label{fig8}
\end{figure}

In Figs.(\ref{fig6})-(\ref{fig8}), we investigate the variation of DPT properties of the kinetic Ising model in the presence of a weak white noise such as $\sigma=0.1$. In Fig. \ref{fig6}, in order to compare the dynamics of the system, 
we plot  the time series of the magnetization $m(t)$ as a function of time (in terms of MCSS)  for $\sigma=0.0$ (without-noise) and  $\sigma=0.1$ (without-noise) cases, respectively. It is clear from Fig. \ref{fig6} that 
the noisy input ($h(t)$) causes a noisy response ($m(t)$) in the system.

In Fig. \ref{fig7}, we further analyze the influence of $\sigma$ on the time series of the order parameter $Q$ (Eq. \ref{eq7}), as well as on the half-period $t_{1/2}$ dependence of dynamic quantities $\langle |Q|\rangle_{L}$ and $\chi^{Q}_{L}$.
From Fig. \ref{fig7}a, it seems that a DPT in the presence of noisy input field $(\sigma=0.1)$ may take place between dynamically ordered and dynamically disordered phases. According to Figs. \ref{fig7}b and \ref{fig7}c, 
$\langle |Q|\rangle_{L}$ evolves from saturation  value to zero with increasing $t_{1/2}$ whereas $\chi^{Q}_{L}$ exhibits a divergent behavior around the critical half period value. 

In order to clarify whether the results presented in Fig. \ref{fig7} indicate a true dynamic phase transition or not, we have performed finite size scaling analysis given by Eqs. (\ref{eq11}) and (\ref{eq12}) 
when the noise parameter is selected as  $\sigma=0.1$. The results are given in Fig. \ref{fig8}. From this figure, we can observe that $\chi^{Q}_{L}$ curves exhibit size dependent maximum, i.e. a divergent behavior around $t_{1/2}^{c}$. 
Regarding the Binder cumulant analysis, as a consequence of possible correction-to-scaling effects \cite{sides3}, and the large fluctuations due to the presence of noisy magnetic field,
we can not identify a distinct intersection point $U_{L}^{*}$ for different $L$. Hence, we have estimated the location of the intersection
by examining and crossing the successive couples of $U_{L}$ curves with different $L$ values such as $U_{256}$ with $U_{180}$, and $U_{180}$ with $U_{128}$, and so on. Averaging over the obtained values, we roughly estimate
the location of critical half period as $t_{1/2}^{c}=73$. Using this value, as well as the value obtained from the peak value of $\chi_{L}^{Q}$, we estimate the critical exponents as $\beta/\nu=0.047$, $\gamma/\nu(t_{1/2}^{c})=2.201$, and 
$\gamma/\nu(peak)=2.196$ for the kinetic Ising model in the presence of white noise with $\sigma=0.1$. These results are displayed in Fig. \ref{fig8}. Since the obtained critical exponent values are definitely different from those obtained
for clean system (i.e. in the absence of noise), we can conclude that even in the presence of weak randomness, DPT observed in the system does not fall
into a universality class of the conventional kinetic (and also equilibrium) Ising model. 

\section{Conclusions}\label{conclude}
In conclusion, we have performed Monte Carlo simulations for the investigation of dynamic phase transition properties of 2D kinetic Ising model in the presence of additive white noise. As a starting point of our discussions,
we have investigated the equilibrium properties $(h_{0}=0.0)$ in the presence of noisy magnetic field, and we found that 
equilibrium Ising model in the presence of additive white noise does not exhibit conventional order-disorder transitions with increasing randomness.
Regarding the dynamic phase transition behavior, we performed finite size scaling analysis in the absence of noise, and we obtained the critical period $P_{c}=184$, and critical exponents $\beta/\nu=0.134$, and $\gamma/\nu=1.749$. 
These results mean that the DPT of kinetic Ising model driven by a sinusoidally oscillating magnetic field in the absence of white noise is in the same universality class with its equilibrium counterpart

On the other hand, in the presence of a weak noise such as $\sigma=0.1$, the obtained critical exponents $\beta/\nu=0.047$, $\gamma/\nu(t_{1/2}^{c})=2.201$, and 
$\gamma/\nu(peak)=2.196$ indicate that the DPT observed in the system does not fall
into a universality class of the conventional dynamic (and also equilibrium) universality class of the Ising model. 

We hope that the results presented in this work stimulates further interest in research of the DPT in stochastic systems.
\section*{Acknowledgements}
The numerical calculations reported in this paper were performed at TUBITAK ULAKBIM High Performance and Grid Computing Center (TR-Grid e-Infrastructure).


\end{document}